\documentclass[preprint,superscriptaddress,amsmath,amssymb,aps,floatfix,nofootinbib,longbibliography]{revtex4-1}

\usepackage[pdftex]{graphicx}
\usepackage{bm}
\usepackage{color}
\usepackage{times}
\usepackage{amsmath}
\usepackage{amssymb}
\usepackage{mathtools}
\usepackage{setspace}

\begin{document}

\title{Gas injection and leakage in layered aquifers}

\author{Luke T. Jenkins}
\affiliation{Department of Earth Sciences, University of Oxford, Oxford, OX1 3AN, UK}
\affiliation{Department of Engineering Science, University of Oxford, Oxford, OX1 3PJ, UK}

\author{Martino Foschi}
\affiliation{Department of Earth Sciences, University of Oxford, Oxford, OX1 3AN, UK}

\author{Christopher W. MacMinn}
\email{christopher.macminn@eng.ox.ac.uk}
\affiliation{Department of Engineering Science, University of Oxford, Oxford, OX1 3PJ, UK}

\date{\today}


\begin{abstract}
Carbon dioxide (CO$_2$) injection into saline aquifers is one method of mitigating anthropogenic climate change. To ensure secure storage of this CO$_2$, it is important to understand the interaction of CO$_2$ injection and migration with geological layering. For example, seismic monitoring at the Sleipner pilot project suggests that the injected CO$_2$ is ponding against, and leaking across, a series of thin, intermediate seals. Here, we develop a gravity-current model for weakly compressible, two-phase fluid migration in a system of layered aquifers. Our model includes vertical leakage of both water and gas across seals, where the latter is subject to a capillary entry threshold. We demonstrate that the buildup of capillary pressure is very sensitive to the conductivity and connectivity of water films in the gas region. We identify two associated limiting cases, where gas obstructs water flow either completely or not at all. We then explore the parameters that govern gas leakage and the resulting fluid distributions---demonstrating that this problem involves a complex interplay between pressure dissipation, capillary pressure buildup, and fluid migration. We show that decreasing the relative permeability of water in the gas region can initiate gas leakage or significantly increase the amount of gas leakage. Finally, we apply our model to rock properties expected for Sleipner and show that CO$_2$ injection may build up sufficient capillary pressure to invade the seals---suggesting that, contrary to conventional wisdom, CO$_2$ may be able to leak across the intermediate seals at Sleipner in the absence of a focused conduit.
\end{abstract}

\maketitle

\section{Introduction}

Carbon capture and geological storage (CCS) involves injecting large amounts of carbon dioxide (CO$_2$) into saline aquifers for long-term storage. In order to achieve a meaningful reduction in CO$_2$ emissions to the atmosphere, CCS would need to be implemented at a very large scale~\citep[\textit{e.g.},][]{ipcc-cambridge-2005}. A key factor in the design and implementation of CCS is that of \textit{storage security}: The injected CO$_2$ should remain safely within the target aquifer. Upward leakage of CO$_2$ into overlying layers could have undesirable environmental consequences, such as the contamination of drinking water~\citep[\textit{e.g.},][]{west2005issue, little2010potential}.

The Sleipner CCS pilot project is the largest and longest-running test of CO$_2$ injection for dedicated storage, involving CO$_2$ injection into a sandstone saline aquifer (Utsira Formation, North Sea) at an average rate of about $1\,\mathrm{Mt}
\,\mathrm{y}^{-1}$ since 1996. Saline aquifers are layers of rock with relatively high permeability, such as sandstone, that are bounded above and below by sealing layers (``seals'') with much lower permeability, such as shale or mudstone. During CO$_2$ injection into such aquifers, the buoyant and relatively low-viscosity CO$_2$ will tend to rise, spread, and migrate as a coherent plume known as a gravity current~\citep[\textit{e.g.},][]{huppert-jfm-1995, nordbotten2006similarity, hesse-jfm-2007, gasda-compgeosci-2009, juanes-tpm-2010}. As a result, the primary barrier to leakage of CO$_2$ out of the target aquifer is the integrity of the aquifer's caprock, which is the seal that forms its uppermost boundary. For a competent seal, gas leakage is blocked by a large capillary entry pressure due to the fine-grained microstructure of the rock.

The injection and subsequent migration of the CO$_2$ at Sleipner has been monitored via periodic seismic surveys. The resulting images show a striking and unanticipated example of CO$_2$ leakage, revealing as many as nine separate CO$_2$ plumes stratified across a sequence of thin mudstone seals interbedded with the sandstone of the aquifer~\citep{zweigel2004geology, boait-jgr-2012}. These mudstone layers are too thin to be resolved in seismic images, so their lateral extent and the precise mechanism by which the CO$_2$ is able to leak across them remains unclear. If the layers lack lateral continuity, then the CO$_2$ could be spilling over their lateral edges~\citep[\textit{e.g.},][]{hesse-grl-2010}. Alternatively,
the CO$_2$ could be flowing across these layers via a focused conduit such as a cross-cutting fault or fracture, or it could be flowing through the pore space of the layers themselves~\citep[\textit{e.g.},][]{foschi-aapg-2018}. The latter is only possible if pressure in the gas exceeds the capillary entry pressure of the seal via some combination of buoyancy and injection pressure.

Leakage across a seal via focused conduits (`focused leakage') has attracted particular attention, and has been studied from a variety of perspectives. For example, several studies have considered the impact of leakage through a fault or fracture on the migration of a buoyant plume along the top seal (or a dense plume along the bottom seal) of a semi-infinite porous layer~\citep[\textit{e.g.},][]{PritchardFracture, NeufeldFissure, NeufeldVellaleak1, NeufeldVellaleak2}. Leakage through a seal by overcoming its capillary entry pressure (`distributed leakage') has attracted less attention. \citet{pritchard-jfm-2001}, \citet{ActonDrainage}, and \citet{farcas-jfm-2009} all studied the impact of distributed leakage on the migration of an unconfined plume (\textit{i.e.}, neglecting the ambient fluid); all of these studies neglected capillary entry pressure, such that the plume always leaks and the local leakage rate is proportional to the local plume thickness. \citet{woods2009capillary} introduced capillary entry pressure to this problem, such that there is a minimum local plume thickness for which leakage will occur.

In leakage from an unconfined porous layer, the ambient fluid is irrelevant to the problem and the migration and leakage of the plume are driven strictly by buoyancy (see references above). In the confined version of this problem, where the porous layer has a finite thickness, migration and leakage of the plume become strongly coupled to migration and leakage of the ambient fluid~\citep{gunn-jfm-2011, kang-wrr-2014, PeglerMigration, pegler-jfm-2015}. Of the studies mentioned above on both focused and distributed leakage, only \citet{PeglerMigration} considered what happens to the fluids after leaking; they assumed that leaked fluids were injected into an overlying aquifer where they would continue to migrate due to pressure and buoyancy, such that the subsequent rate of leakage was then coupled with flow in the overlying aquifer.

In a confined layer, distributed leakage is complicated by the fact that both fluids can cross the seals at any point, leading to a rich flow problem that has not previously been explored. \citet{jenkins-jfm-2019} recently studied distributed leakage of the ambient fluid (brine) during gas injection into a layered system of horizontal aquifers separated by thin seals, but assuming that the gas cannot leak (\textit{i.e.}, that the entry pressure is never exceeded). Distributed brine leakage is interesting and important in its own right, because it enables strong vertical pressure dissipation and therefore plays an important role in mitigating pressure buildup during gas injection~\citep{birkholzer2009large, nicot-ijggc-2008, ChangHesse2013}. \citet{jenkins-jfm-2019} showed that vertical pressure dissipation is also coupled to the motion of the gas plume, leading to a more compact plume shape by suppressing the formation of the advancing gas tongue. Here, we extend the work of \citet{jenkins-jfm-2019} to account for distributed gas leakage.

A key aspect of the present study is that, as in \citet{jenkins-jfm-2019}, we focus on a layered system. This enables a strong coupling between injection, migration, leakage, and pressure dissipation across layers. For example, gas injection drives brine leakage by pressurising the target aquifer relative to the over- and underlying aquifers, but this brine leakage acts to dissipate the injection pressure into the over- and underlying aquifers and thereby imposes a negative feedback on further leakage. We show here that this coupling becomes even more complex when gas leakage is introduced. In \S\ref{s:model}, we review the model of \citet{jenkins-jfm-2019} and extend it to include gas leakage. In \S\ref{s:results}, we outline our numerical scheme and then explore the different factors that control the buildup of capillary pressure along the underside of the seal. We then explore the parameter space that governs gas leakage. In \S\ref{s:discussion}, we discuss the implications of our results for CCS, with a particular focus on CO$_2$ injection at Sleipner. Note that we focus here on the likelihood of gas leakage, and on the impact of gas leakage on the spatial distribution of gas at the end of injection. We plan to study the time evolution of the gas plume and the pressure field, both during and after injection, in future work.

\section{Theoretical model}\label{s:model}

\begin{figure}
    \centering
    \includegraphics[width=0.6\textwidth]{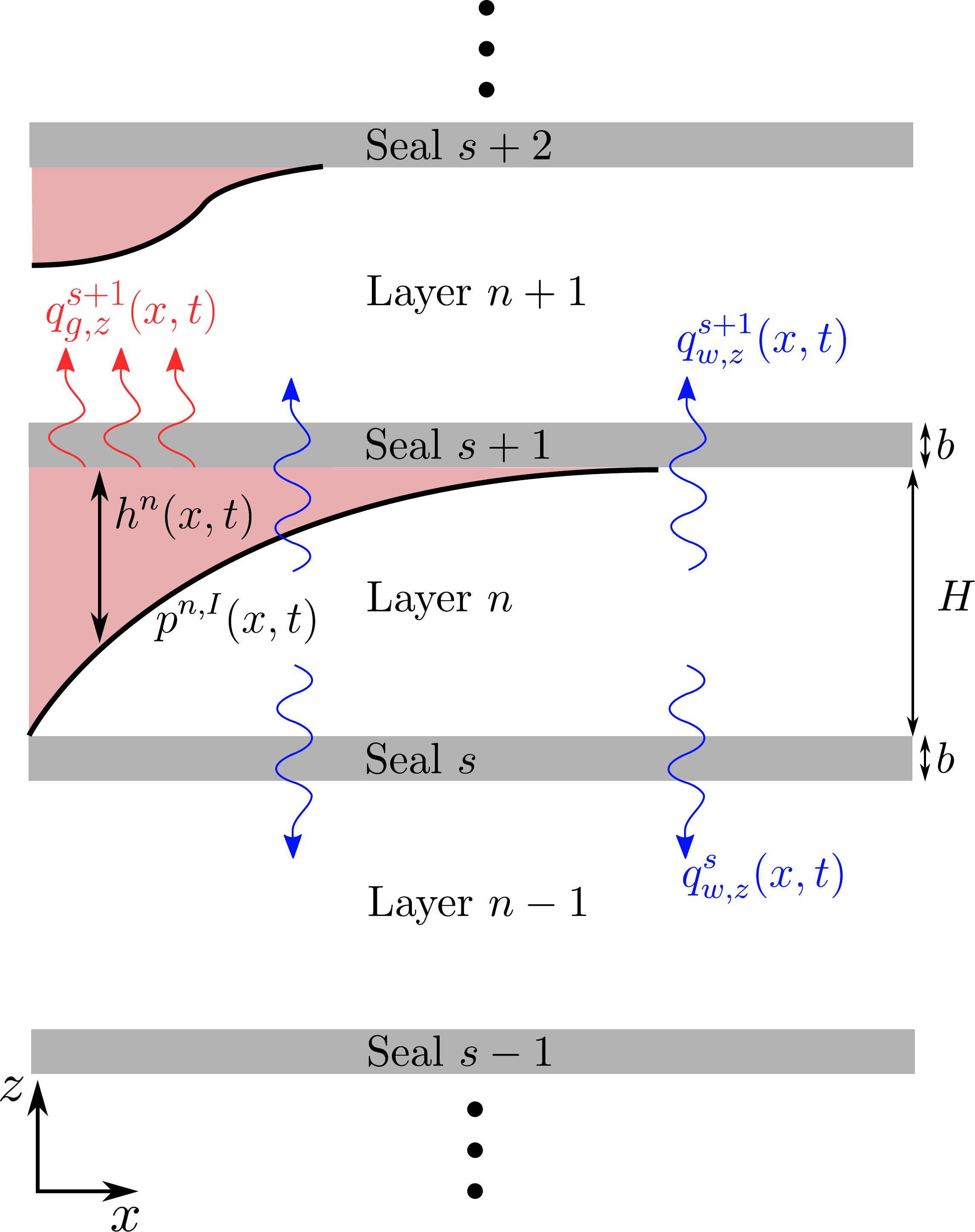}
    \caption{A section of our model system, which comprises a sequence of aquifers of thickness $H$ alternating with seals of thickness $b$. The gas-saturated region is shown in red.   \label{fig:gas_leak_diagram} }
\end{figure}

Our geological setting is the same as in \citet{jenkins-jfm-2019}. We consider a layered system comprising an alternating stack of $N_z$ aquifers and $N_z+1$ seals (Figure~\ref{fig:gas_leak_diagram}). Each aquifer is bounded above and below by a seal, and we count the aquifers and seals from the bottom up, such that the deepest and shallowest aquifers are 1 and $N_z$ respectively, and the deepest and shallowest seals are 1 and $N_z+1$ respectively.

We assume that all aquifers have the same uniform thickness $H$, porosity $\phi$, and isotropic permeability $k$, and that all seals have the same uniform thickness $b$ $(b\ll{}H)$ and isotropic permeability $k_s$ $(k_s\ll{}k)$. It would be straightforward to extend our model to include anisotropy, variations in layer properties, or lateral heterogeneity, but we omit these complications here for simplicity.

We consider two immiscible phases of different density: A buoyant, non-wetting phase (`gas') and a denser, wetting ambient phase (`water'). In the context of CCS, the former would be supercritical CO$_2$ (not strictly a gas) and the latter would be saline groundwater, also commonly referred to as brine. We denote phase identity with a subscript $\alpha$, with $\alpha=w$ for water and $\alpha=g$ for gas. We account for the weak compressibility of both fluids by allowing their densities $\rho_\alpha$ to vary linearly with pressure about a reference state,
\begin{equation}\label{eq:comp}
    \rho_\alpha(p) =\rho_\alpha^0\left[1+c_\alpha(p_\alpha-p^0)\right],
\end{equation}
where $p_\alpha$ is the pressure of phase $\alpha$, $\rho_\alpha^0$ is the density of phase $\alpha$ at reference pressure $p^0$, and $c_\alpha$ is the compressibility of phase $\alpha$ about $p^0$ ($c_\alpha\equiv(1/\rho_{\alpha}^0)(\mathrm{d}\rho_\alpha/\mathrm{d}p)|_{p^0}$). For pressures typically associated with natural fluid migration and subsurface engineering, we expect that $c_w(p_w-p^0)\ll{}1$. We therefore assume that $\rho_w\approx{}\rho_{w}^{0}$, which simplifies our analysis below. We do not make this assumption for gas since we expect that $c_g\gg{}c_w$.

As in \citet{jenkins-jfm-2019}, we allow for vertical and lateral pressure dissipation via brine flow through the aquifers and across the seals. Unlike in \citet{jenkins-jfm-2019}, we now also allow for gas flow across the seals, subject to an appropriate capillary entry threshold. The central goal of this study is to explore the coupling between pressure dissipation and gas leakage during injection.

We focus on flow in the $x$--$z$ plane, where gravity points in the negative $z$ direction. We assume symmetry along the $y$ direction (into the page). We denote the top and bottom of aquifer $n$ by $z^{n,T}$ and $z^{n,B}$, respectively, such that $z^{n,T}-z^{n,B}\equiv{}H$ and $z^{n+1,B}-z^{n,T}\equiv{}b$ for all $n$ (Figure~\ref{fig:gas_leak_diagram}).

\subsection{Flow in aquifer $n$}

The theoretical framework for our model and the majority of the underlying assumptions are the same as those in \citet{jenkins-jfm-2019}. As a result, we proceed by outlining the key points and highlighting the differences from that previous study. We again invoke a two-phase gravity-current formulation in which the fluids are strongly segregated by gravity --- that is, we study the flow of a coherent plume of gas relative to ambient water.

\subsubsection{Water in aquifer $n$}\label{sec:derivation-water}

We now outline the derivation of the governing partial differential equation (PDE) for water in aquifer $n$, which is identical to that in \citet{jenkins-jfm-2019}. Conservation of mass for the water in aquifer $n$ is given by
\begin{equation}\label{eq:consW}
    \frac{\partial}{\partial{t}}\left(\rho_w{}\phi{}s_w\right) +\boldsymbol{\nabla}\cdot\left(\rho_w\boldsymbol{q}_w\right) = \mathcal{I}_w,
\end{equation}
where $s_w$ is the water saturation, $\boldsymbol{q}_w$ is the Darcy flux of water, and $\mathcal{I}_w$ is the local mass rate of water injection per unit volume. The Darcy flux of water is given by Darcy's law,
\begin{equation}\label{eq:DarcyW}
    \boldsymbol{q}_w =-\frac{kk_{rw}}{\mu_w} \,\left(\boldsymbol{\nabla}p_w+\rho_w{}g\hat{\boldsymbol{e}}_z\right),
\end{equation}
where $k_{rw}$ is the relative permeability of water, $\mu_w$ is the dynamic viscosity of water, which we take to be constant and uniform, $p_w$ is the water pressure, $g$ is body force per unit mass due to gravity, and $\hat{\boldsymbol{e}}_z$ is the unit vector in the positive $z$ direction.

The vertical component of Darcy's law (Eq.~\ref{eq:DarcyW}) can be rearranged to give an expression for the vertical pressure gradient,
\begin{equation}\label{eq:pw_from_Darcy}
    \frac{\partial{p_w}}{\partial{z}}=-\rho_w{}g-\frac{q_{w,z}}{\lambda_w},
\end{equation}
where $q_{w,z}$ is the vertical component of the water flux and $\lambda_w\equiv{}kk_{rw}/\mu_w$ is the water mobility. The classical model for a gravity current in an aquifer with impermeable seals involves assuming that the fluids are in `vertical equilibrium', meaning that the flow is predominantly horizontal and the vertical pressure distribution is therefore nearly hydrostatic (\textit{i.e.}, $q_{w,z}\ll{}\rho_wg\lambda_w$ $\implies$ $\partial{p_w}/ \partial{z}\approx-\rho_wg$) \citep[\textit{e.g.},][]{Bear1972,huppert-jfm-1995}. This assumption allows for direct calculation of $p_w$ by integrating Equation~\eqref{eq:pw_from_Darcy}, with the resulting expression for $p_w$ being linear in $z$. \citet{jenkins-jfm-2019} extended this concept to allow for weak vertical flow of water by instead assuming that $q_{w,z}$ has a simple, continuous and piecewise-linear structure in $z$, as originally suggested by \citet{nordbotten2006improved} for flow near a well. The expression is
\begin{equation}\label{eq:qwzL2}
    q_{w,z}(x,z,t)\approx
    \begin{cases}
        q_{w,z}^{n,B} + \displaystyle\left(\frac{z-z^{n,B}}{z^{n,I}-z^{n,B}}\right)(q_{w,z}^{n,T}-q_{w,z}^{n,B}) \quad &z^{n,B}\leq{}z<z^{n,I}, \\[1em]
        q_{w,z}^{n,T} \quad &z^{n,I}\leq{}z\leq{}z^{n,T},
    \end{cases}
\end{equation}
where $q_{w,z}^{n,B}(x,t)$ and $q_{w,z}^{n,T}(x,t)$ are the vertical fluxes of water through the lower and upper seals of aquifer $n$, respectively. Thus, $q_{w,z}$ varies linearly from $q_{w,z}^{n,B}$ at the bottom seal to $q_{w,z}^{n,T}$ at the gas-water interface, and is then uniform and equal to $q_{w,z}^{n,T}$ from the gas-water interface to the top seal. This approach still allows for direct calculation of $p_w$ by integrating Equation~\eqref{eq:pw_from_Darcy}, with the resulting expression for $p_w$ being continuous and piecewise-parabolic in $z$. Note that the fluxes $q_{w,z}^{n,B}$ and $q_{w,z}^{n,T}$ are unknown, and will be determined through global conservation of mass. Combining this result with the horizontal component of Equation~\eqref{eq:DarcyW}, and then substituting into Equation~\eqref{eq:consW} and integrating over the full thickness of the aquifer, we arrive at the governing equation for water:
\begin{equation}\label{eq:W_PDE}
    \begin{split}
        \phi\bigg[(H-s_g{}h^n)(c_r&+c_w)\frac{\partial{p^n}}{\partial{t}}
-s_g\frac{\partial{h^n}}{\partial{t}}\bigg] -\frac{\partial}{\partial{x}}\bigg\{\lambda_w(H-h^n)\bigg[\frac{\partial{p^n}}{\partial{x}} -\rho_{w}g\frac{\partial{h^n}}{\partial{x}}\bigg] \\
    &+\frac{1}{6}\frac{\partial}{\partial{x}}\bigg[(H-h^n)^2(q_{w,z}^{n,B}+2q_{w,z}^{n,T})\bigg]\bigg\} = -(q_{w,z}^{n,T}-q_{w,z}^{n,B}) + \frac{\mathcal{I}_w^{n}H}{\rho_w},
    \end{split}
\end{equation}
where $c_r$ is the rock compressibility, $h^n$ is the thickness of the gas layer in aquifer $n$, $p^n$ is the water pressure along the gas-water interface in aquifer $n$, and $\mathcal{I}_w^n$ is the vertically averaged mass injection rate of water per unit volume into aquifer $n$ (Figure~\ref{fig:gas_leak_diagram}). Recall that we have assumed throughout that $\rho_w\approx{}\rho_w^0$. We refer the reader to \citet{jenkins-jfm-2019} for more details and discussion related to this derivation.

\subsubsection{Gas in aquifer $n$}\label{sec:derivation-gas}

Conservation of mass for gas in aquifer $n$ is given by
\begin{equation}\label{eq:consG}
    \frac{\partial}{\partial{t}}\left(\rho_g{}\phi{}s_g\right) +\boldsymbol{\nabla}\cdot\left(\rho_g\boldsymbol{q}_g\right) = \mathcal{I}_g,
\end{equation}
where $s_g$ is the saturation of gas, $\boldsymbol{q}_g$ is the Darcy flux of gas, and $\mathcal{I}_g$ is the local mass rate of gas injection per unit volume. The Darcy flux of gas is given by Darcy's law,
\begin{equation}\label{eq:DarcyG}
    \boldsymbol{q}_g =-\frac{kk_{rg}}{\mu_g}\,\left(\boldsymbol{\nabla}p_g+\rho_g{}g\hat{\boldsymbol{e}}_z\right),
\end{equation}
where $k_{rg}$ is the relative permeability of gas, $\mu_g$ is the dynamic viscosity of gas, which we take to be constant and uniform, and $p_g$ is the gas pressure.

The vertical component of Darcy's law (Eq.~\ref{eq:DarcyG}) can again be rearranged to give an expression for the vertical pressure gradient,
\begin{equation}\label{eq:pg_from_Darcy}
    \frac{\partial{p_g}}{\partial{z}}=-\rho_gg-\frac{q_{g,z}}{\lambda_g},
\end{equation}
where $q_{g,z}$ is the vertical component of the gas flux and $\lambda_g\equiv{}kk_{rg}/\mu_g$ is the mobility of the gas. As for water, the classical approach is to calculate $p_g$ by assuming vertical equilibrium in the gas (\textit{i.e.}, $q_{g,z}\ll{}\rho_gg\lambda_g \implies \partial{p_g}/\partial{z}\approx{}-\rho_gg$) and then integrating in $z$ \citep[\textit{e.g.},][]{Bear1972,huppert-jfm-1995}. \citet{jenkins-jfm-2019} made the same assumption, motivated by the scenario where the capillary entry pressure of the seals, $p_c^E$, was large enough to prevent gas from entering. We now relax this assumption, allowing for gas leakage through the seals when this entry pressure is exceeded by assuming a weak vertical flow of gas in the gas region. We assume that this vertical gas flux $q_{g,z}$ is uniform in $z$,
\begin{equation}\label{eq:qgzL2}
    q_{g,z}(x,z,t) \approx q_{g,z}^{n,T}\quad z^{n,I}\leq{}z<z^{n,T},
\end{equation}
where $q_{g,z}^{n,T}(x,t)$ is the vertical flux of gas through the upper seal of aquifer $n$.

Substituting Eq.~\eqref{eq:qgzL2} into Eq.~\eqref{eq:pg_from_Darcy} and integrating, we arrive at
\begin{equation}\label{eq:p_g}
    p_g^n(x,z,t) = p^n - \bigg(\rho_g^{n}g + \frac{q_{g,z}^{n,T}}{\lambda_g}\bigg)(z-z^{n,I}),
\end{equation}
where $\rho_g^n$ is the vertically averaged gas density in aquifer $n$. Note that we have neglected the capillary pressure at the gas-water interface (\textit{i.e.}, the entry pressure of the aquifer) relative to hydrostatic varations. Note also that we have neglected terms of size $\rho_{g}^{0}gHc_g\ll{}1$ in this calculation, taking the gas density to be approximately vertically uniform within the aquifer.

Substituting Eq.~\eqref{eq:p_g} into the horizontal component of Eq.~\eqref{eq:DarcyG} and continuing to neglect terms of order $\rho_{g,0}gHc_g\ll{}1$, we arrive at an expression for the horizontal flux of gas in aquifer $n$,
\begin{equation}
    q_{g,x}^n = -\lambda_g\bigg[\frac{\partial{p_g^n}}{\partial x} -\bigg(\rho_g^{n}g+\frac{q_{g,z}^{n,T}}{\lambda_g}\bigg)\frac{\partial{h^n}}{\partial{x}} -\frac{1}{\lambda_g}\frac{\partial{q_{g,z}^{n,T}}}{\partial{x}}\left(z-z^{n,I}\right)\bigg],
\end{equation}
where we have used the fact that $z^{n,I}=z^{n,T}-h^n$.

We now integrate Eq.~\eqref{eq:consG} over the thickness of the aquifer,
\begin{equation}\label{eq:consG_VI}
    \int_{z_n^B}^{z_n^T}\,\frac{\partial}{\partial{t}}\left(\rho_g{}s_g\phi\right)\,\mathrm{d}z+\int_{z_n^B}^{z_n^T}\,\boldsymbol{\nabla}\cdot\left(\rho_g\boldsymbol{q}_g\right)\,\mathrm{d}z = \int_{z_n^B}^{z_n^T}\,\mathcal{I}_g\,\mathrm{d}z.
\end{equation}
As in \citet{jenkins-jfm-2019}, the first term of Eq.~\eqref{eq:consG_VI} becomes
\begin{equation}
    \begin{split}\label{eq:G_T1}
        \int_{z_n^B}^{z_n^T}\,\frac{\partial}{\partial{t}}(\rho_g{}\phi{}s_g)\,\mathrm{d}z &\approx{}\frac{\partial}{\partial{t}}(\rho_g^n\phi{}s_gh^n) \\
        &\approx\rho_g^n\phi{}s_g\bigg[(c_r+  \frac{\rho_g^0}{\rho_g^n}c_g)h^n\frac{\partial{p^n}}{\partial{t}} + \frac{\partial{h^n}}{\partial{t}} \bigg],
    \end{split}
\end{equation}
where $s_g$ is now the constant and uniform saturation of gas in the gas region and $c_r\equiv(1/\phi)(\partial{\phi}/\partial{p})$ is the rock compressibility \citep[\textit{e.g.},][]{Bear1972}. The density ratio $\rho_g^0/\rho_g^n$ is often approximated as unity; we avoid this approximation to ensure conservation of mass when the gas phase is moderately compressible, as would be the case for methane. The second term of Eq.~\eqref{eq:consG_VI} becomes
\begin{equation}\label{eq:G_T2}
    \begin{split}
        \int_{z^{n,B}}^{z^{n,T}}\,\boldsymbol{\nabla}&\cdot\left(\rho_g\boldsymbol{q}_g\right)\,\mathrm{d}z =\frac{\partial}{\partial{x}}\left(\int_{z^{n,B}}^{z^{n,T}}\,\rho_gq_{g,x}\,\mathrm{d}z\right)+(\rho_g q_{g,z})\Big|_{z^{n,B}}^{z^{n,T}} \\
        \approx&\frac{\partial}{\partial{x}}\bigg\{-\rho_{g}^{n}\lambda_{g}h^{n}\left[\frac{\partial{p^{n}}}{\partial{x}} -\rho_{g}^{n}g\frac{\partial{h^n}}{\partial{x}}\right] +\rho_{g}^{n}\frac{\partial}{\partial{x}}\left[\frac{1}{2}(h^n)^2q_{g,z}^{n,T}\right]\bigg\} \\
        &+\left(\rho_g^nq_{g,z}^{n,T}-\rho_g^nq_{g,z}^{n,B}\right),
    \end{split}
\end{equation}
Recombining Eqs.~\eqref{eq:G_T1} and \eqref{eq:G_T2} with Eq.~\eqref{eq:consG_VI} yields our governing equation for a compressible buoyant gravity current of gas with additional terms related to injection, weak vertical gas flow, and gas leakage,
\begin{equation} \label{eq:G_PDE}
    \begin{split}
        \rho_g^n\phi{}s_g\bigg[\bigg(c_r+\frac{\rho_g^0}{\rho_g^n}c_g\bigg)&h^n\frac{\partial{p^n}}{\partial{t}} +\frac{\partial{h^n}}{\partial{t}}\bigg] -\frac{\partial}{\partial{x}}\bigg\{\rho_g^{n}\lambda_{g}h^{n}\left[\frac{\partial{p^n}}{\partial{x}}-\rho_{g}^{n}g\frac{\partial{h}^n}{\partial{x}}\right] \\
        &+\rho_g^n\frac{\partial}{\partial{x}}\left[\frac{1}{2}(h^n)^2\,q_{g,z}^{n,T}\right]\bigg\} =-\left(\rho_g^{n}q_{g,z}^{n,T}-\rho_g^{n}q_{g,z}^{n,B}\right) +\mathcal{I}_{g}^{n}H,
    \end{split}
\end{equation}
where $\mathcal{I}_g^n$ is the vertically averaged mass injection rate of gas per unit volume into aquifer $n$. Equations~\eqref{eq:W_PDE} and \eqref{eq:G_PDE} are two coupled nonlinear partial differential equations (PDEs) in $h^n$ and $p^n$. To close this system, we need expressions for the vertical fluxes of water ($q_{w,z}^{n,B}$ and $q_{w,z}^{n,T}$) and gas ($q_{g,z}^{n,B}$ and $q_{g,z}^{n,T}$). Note that, because we have assumed that there is no gas in the seals or the water regions of the aquifers, gas that leaks upward out of one aquifer appears immediately in the gas plume in the aquifer above --- in other words, we neglect the transit time between one gas plume and the next.

\subsection{Coupling the aquifers with vertical fluxes}\label{sec:derivation-fluxes}

\subsubsection{Vertical water fluxes}

For the vertical fluxes of water across the seals, our approach and results are identical to those of \citet{jenkins-jfm-2019}. We assume that there is horizontal flow and no storage within the seals, such that the mass flux of water entering seal $s$ from aquifer $n-1$ must equal the mass flux of water exiting seal $s$ into aquifer $n$:
\begin{equation}
    \rho_w^{n-1}q_{w,z}^{n-1,T}=\rho_w^{n}q_{w,z}^{n,B}=\rho_w^{s}q_{w,z}^s,
\end{equation}
where $\rho_w^{s}$ is the density of water in seal $s$ and $q_{w,z}^s$ is the flux of water through seal $s$. Recall that we assume that $\rho_w\approx{}\rho_w^0$ throughout the system, such that $\rho_w^{n-1}\approx{}\rho_w^n\approx{}\rho_w^s\approx{}\rho_w^0$. We the calculate the vertical flux $q_{w,z}^s$ via Darcy's law, introducing expressions for the pressures at the top and bottom of each aquifer from our analysis above. The result can be written
\begin{equation}\label{eq:qs_system}
    \begin{split}
        \left(\frac{H-h^n}{2\lambda_w}\right)q_{w,z}^{s+1} +\bigg(\frac{h^{n-1}}{\lambda_w^\star}+\frac{b}{\lambda_w^s}&+\frac{H-h^n}{2\lambda_w}\bigg)q_{w,z}^s \\
        &=-\bigg[p^n-p^{n-1}+\rho_{w}^{0}g(h^{n-1}+b+H-h^n)\bigg],
    \end{split}
\end{equation}
where $\lambda_w^\star=kk_{rw}^\star/\mu_w$ is the mobility of water in the gas regions of the aquifers, with $k_{rw}^\star$ the relative permeability of water in those regions, and $\lambda_w^s=k_s/\mu_w$ is the mobility of water in the seals. Equation~\eqref{eq:qs_system} is a linear system of $N_z-1$ coupled algebraic equations in the $N_z-1$ unknown fluxes $q_{w,z}^s$ for $s=2\ldots{}N_z-1$, from which we can solve for $q_{w,z}^s$ in terms of $p^n$ and $h^n$. Recall that we take the bottom-most and top-most seals to be impermeable, such that $q_{w,z}^1=q_{w,z}^{N_z+1}=0$.

\subsubsection{Vertical gas fluxes}

For the vertical fluxes of gas across the seals, we follow a similar procedure to that for water. With no horizontal flow and no storage within the seals, the mass flux of gas entering seal $s$ from aquifer $n-1$ must equal the mass flux of gas exiting seal $s$ into aquifer $n$:
\begin{equation}
    \rho_g^{n-1}q_{g,z}^{n-1,T}=\rho_g^{n}q_{g,z}^{n,B}=\rho_g^{s}q_{g,z}^s,
\end{equation}
where $\rho_g^{s}$ is the density of gas in seal $s$ and $q_{g,z}^s$ is the flux of gas through seal $s$. Unlike the water, we allow the gas to be moderately compressible; as a result, the three gas densities  $\rho_g^{n-1}$, $\rho_g^n$, and $\rho_g^s$ are not necessarily equal. We calculate the former two densities from the associated pressures $p^{n-1}$ and $p^n$ via Eq.~\eqref{eq:comp}, and we take the latter to be the average of the former two,
\begin{equation}
    \rho_g^{s}=\frac{1}{2}(\rho_g^{n-1}+\rho_g^{n}).
\end{equation}

We next write Darcy's law for $q_{g,z}^s$, including a capillary threshold condition that prevents gas leakage unless the capillary pressure along the underside of seal $s$ exceeds the associated entry pressure $p_c^E$,
\begin{equation}\label{eq:DarcyGzs-step}
    q_{g,z}^{s}(x,t)=
    \begin{cases}
        -\lambda_g^{s}\displaystyle\bigg(\frac{p_g^{n,B}-p_g^{n-1,T}}{b}+\rho_g^{s}\bigg) \quad &p_c^{n-1,T}>p_c^E, \\[1em]
        0 \quad &p_c^{n-1,T}<p_c^E,
    \end{cases}
\end{equation}
where $p_c^{n-1,T}=p_g^{n-1,T}-p_w^{n-1,T}$ is the capillary pressure at the top of aquifer $n-1$ (bottom of seal $s$). We smooth the sharp transition across $p_c^E$ by introducing a step-like transition function $\mathcal{R}(p_c^{n-1,T})$ and rewriting Eq.~\eqref{eq:DarcyGzs-step} as
\begin{equation}\label{eq:DarcyGzs-smooth}
    q_{g,z}^{s}(x,t)=-\mathcal{R}\lambda_g^{s}\bigg(\frac{p_g^{n,B}-p_g^{n-1,T}}{b}+\rho_g^{s}\bigg),
\end{equation}
with
\begin{equation}\label{eq:Rpc}
    \mathcal{R}(p_c^{n-1,T})=\frac{1}{2}\left\{1+\tanh\left[\left(\frac{p_c^{n-1,T}-p_c^E}{p_c^E}\right)\vartheta\right]\right\},
\end{equation}
such that $\mathcal{R}\to0$ for $p_c^{n-1,T}<{}p_c^E$ and $\mathcal{R}\to1$ for $p_c^{n-1,T}>{}p_c^E$. The parameter $\vartheta$ controls the sharpness of this transition, with Eq.~\eqref{eq:Rpc} converging to a unit step, and therefore Eq.~\eqref{eq:DarcyGzs-smooth} converging to Eq.~\eqref{eq:DarcyGzs-step}, for $\vartheta\gg{}1$. Note that $\mathcal{R}(p_c^{n,T}=p_c^E)= 1/2$ for all $\vartheta$. Our results should be independent of the particular value of $\vartheta$ as long as it is sufficiently large; we typically use $\vartheta=300$.

Finally, we must relate the pressures $p_g^{n-1,T}$ and $p_g^{n,B}$ to known quantities. Equation~\eqref{eq:p_g} suggests that the former pressure is given by
\begin{equation}\label{eq:pgnT_pre}
    p_g^{n-1,T} = p^{n-1}-\bigg(\rho_g^{n-1}g + \frac{q_{g,z}^{n-1,T}}{\lambda_g}\bigg)h^{n-1}.
\end{equation}
For the latter pressure, we again note that we expect the gas column through seal $s$ and within the water region of aquifer $n$ to be discontinuous. As a result, we expect the gas pressure at the bottom of aquifer $n$ to be controlled by the continuous water column, $p_g^{n,B}\approx{}p_w^{n,B}$, where we again neglect the capillary pressure in the aquifers. This water pressure is itself determined via the separate system of equations for water leakage, Eq.~\eqref{eq:qs_system}. Rearranging Eq.~(2.16) of \citet{jenkins-jfm-2019}, the relevant expression is
\begin{equation}
    p_w^{n,B} = p^n+(H-h^n)\left(\rho_wg+\frac{q_{w,z}^{s+1}+q_{w,z}^s}{2\lambda_w}\right).
\end{equation}
Combining all of these ingredients, we eliminate $q_{g,z}^{n-1,T}$ from Eq.~\eqref{eq:pgnT_pre} and write the result as
\begin{equation}\label{eq:pgnT}
    p_g^{n-1,T} = \frac{1}{\xi_g^{s}}\bigg[p^{n-1}-\rho_g^{n-1}gh^{n-1} +(\xi_g^{s}-1)(p_w^{n,B}+\rho_g^{s}gb)\bigg],
\end{equation}
where the dimensionless quantity $\xi_g^s$ is given by
\begin{equation}\label{eq:xig}
    \xi_g^{s} = 1 + \left(\frac{\rho_g^s\lambda_g^sh^{n-1}}{\rho_g^{n-1}\lambda_gb}\right)\mathcal{R}.
\end{equation}
With Eq.~\eqref{eq:pgnT}, the gas-leakage flux can now be evaluated directly from 
\begin{equation}\label{eq:DarcyGzs}
    q_{g,z}^{s}(x,t)=-\mathcal{R}\lambda_g^{s}\bigg(\frac{p_w^{n,B}-p_g^{n-1,T}}{b}+\rho_g^{s}\bigg),
\end{equation}
where $\mathcal{R}(p_c^{n-1,T})$ is as defined in Eq.~\eqref{eq:Rpc}. Note that Eqs.~\eqref{eq:pgnT}--\eqref{eq:DarcyGzs} reduce to the standard result from vertical equilibrium, $p_g^{n-1,T} =p^{n-1}-\rho_g^{n-1}gh^{n-1}$ and $q_{g,z}^{s}=0$, when the entry pressure is not exceeded ($p_c^{n-1,T}<p_c^E \implies \mathcal{R}=0$). 

To assess the entry-pressure condition and determine the value of $\mathcal{R}$, we must calculate the capillary pressure at the top of each aquifer. Equation~\eqref{eq:pgnT} is an expression for the gas pressure at the top of aquifer $n-1$. Equations~(2.23) and (2.24a) of \citet{jenkins-jfm-2019} can be rearranged to provide an expression for the water pressure at the top of aquifer $n-1$,
\begin{equation}\label{eq:pwnT}
    p_w^{n-1,T} = \frac{1}{\xi_w^{s}}\bigg[p^{n-1}-\rho_wgh^{n-1} +(\xi_w^{s}-1)(p_w^{n,B}+\rho_wgb)\bigg],
\end{equation}
with
\begin{equation}
    \xi_w^{s} = 1+\frac{\lambda_w^{s}h^{n-1}}{\lambda_w^{\star}b}.
\end{equation}
The capillary pressure $p_c^{n-1,T}=p_g^{n-1,T}-p_w^{n-1,T}$ is then given by
\begin{equation}\label{eq:pcNTdim}
    \begin{split}
        p_c^{n-1,T} = &\left(\frac{1}{\xi_g^{s}}-\frac{1}{\xi_w^{s}}\right)p^{n-1} -\left(\frac{\rho_g^{n-1}}{\xi_g^{s}}-\frac{\rho_w}{\xi_w^{s}}\right)gh^{n-1} \\
        &-\left(\frac{1}{\xi_g^{s}}-\frac{1}{\xi_w^{s}}\right)p_w^{n,B} -\left(\frac{\rho_g^{s}}{\xi_g^{s}}-\frac{\rho_w}{\xi_w^{s}}\right)gb-(\rho_w-\rho_g^s)gb.
    \end{split}
\end{equation}
Note that this expression for $p_c^{n-1,T}$ depends on $\mathcal{R}$ via $\xi_g^s$ because the gas pressure at the top of aquifer $n$ depends on whether or not the gas is actively leaking. In principle, Eqs.~\eqref{eq:pgnT}--\eqref{eq:pcNTdim} should be evaluated iteratively to assess the entry-pressure condition and calculate the gas-leakage flux. In practice, $\xi_g^s$ is insensitive to the value of $\mathcal{R}$ because $\lambda_g^s\ll{}\lambda_g$, so it is safe and efficient to assume $\xi_g^s\approx1$ and calculate the other quantities accordingly. Note also that, whereas the connected water column and the resulting continuous water pressure field lead to a system of coupled equations for the $N_z-1$ unknown water leakage fluxes (Eq.~\ref{eq:qs_system}), the disconnected gas column leads to an explicit expression for the each of the $N_z-1$ unknown gas-leakage fluxes (Eq.~\ref{eq:DarcyGzs}), where again $q_{g,z}^1=q_{g,z}^{N_z+1}=0$. Importantly, however, the capillary pressure and the gas-leakage fluxes are strongly coupled to the water pressure field; we explore this coupling in much more detail below.

\subsection{Boundary and initial conditions}

As in \citet{jenkins-jfm-2019}, we consider a system comprised of $N_z$ aquifers alternating with $N_z+1$ seals, all of which extend horizontally from $x=-L_x/2$ to $x=L_x/2$. We assume that the system is initially fully saturated with water (no gas), and that the pressure distribution is initially hydrostatic. We assume the pressure at the lateral boundaries remains hydrostatic for all time and we take the bottom-most and top-most seals to be impermeable ($s=1$ and $s=N_z+1$, respectively). 
For injection of phase $\alpha$ into the horizontal centre of aquifer $n$ at a mass flow rate $\dot{M}_\alpha^n(t)$ per unit length into the page, the appropriate vertically integrated source term is
\begin{equation}
    \mathcal{I}_\alpha^n = \frac{\dot{M}_\alpha^n(t)}{H}\,\delta(x),
\end{equation}
where $\delta(x)$ is the Dirac delta function.

\subsection{Non-dimensionalization}

As in \citet{jenkins-jfm-2019}, we choose characteristic scales based on the injection of gas at a mass flow rate $\dot{M}$ per unit length into the page for a time $\mathcal{T}$. The associated characteristic scales for length, pressure, and vertical flux are
\begin{equation}\label{eq:scales}
    \mathcal{L} \equiv \frac{\dot{M}\mathcal{T}}{2\phi s_g \rho_{g}^{0}H}\,\,, \quad\mathcal{P} \equiv\frac{\phi\mathcal{L}^2}{\lambda_w\mathcal{T}}=\frac{\dot{M}\mathcal{L}}{2\lambda_w s_g \rho_{g}^{0}H}\,\,, \quad\mathrm{and}\quad \mathcal{Q}_z \equiv\frac{\Lambda_w^s \mathcal{P}}{b}.
\end{equation}
The characteristic length $\mathcal{L}$ is the half-width of an incompressible box of gas of mass $\dot{M}\mathcal{T}$ per unit length into the page. The characteristic pressure $\mathcal{P}$ is the pressure drop associated with a Darcy flux $\phi\mathcal{L}/\mathcal{T}$ of water over a distance $\mathcal{L}$. The characteristic vertical flux $\mathcal{Q}_z$ is the vertical flux of water driven by a pressure drop $\mathcal{P}$ across a seal of thickness $b$.

We use the above scales in combination with existing parameters to define the following dimensionless quantities:
\begin{equation}
    \begin{split}
        \tilde{x}\equiv\frac{x}{\mathcal{L}}, \quad
        \tilde{t}\equiv\frac{t}{\mathcal{T}}, \quad
        &\tilde{h}\equiv\frac{h}{H}, \quad \tilde{p}\equiv\frac{p}{\mathcal{P}}, \quad \tilde{q}\equiv\frac{q}{\mathcal{Q}_z}, \\
        &\tilde{b}\equiv\frac{b}{H}, \quad
        \tilde{\rho}_\alpha\equiv\frac{\rho_\alpha}{ \rho_g^0}, \quad
        \mathrm{and}\quad \tilde{\mathcal{I}}_{\alpha}^{n}\equiv\frac{2\mathcal{L}H\,\mathcal{I}_{\alpha}^{n}}{\dot{M}}.
    \end{split}
\end{equation}
We then also introduce the following dimensionless groups:
\begin{subequations}
    \begin{align}
        N_{cw} &\equiv c_w \mathcal{P}\\
        R_{cw} &\equiv c_r/c_w\\
        R_{cf} &\equiv c_g/c_w\\
        R_A &\equiv \mathcal{L}/H\\
        R_d &\equiv \rho_{g}^{0}/\rho_{w}^{0}\\
        N_g &\equiv \rho_{w}^{0}gH/\mathcal{P}\\
        \mathcal{M} &\equiv \lambda_g/(s_g\lambda_w)\\
        \Lambda_w^s &\equiv \lambda_w^sH/(\lambda_w b)\\
        \mathcal{M}_z^s &\equiv \lambda_g^{s}/\lambda_w^{s}\\
        \tilde{p}_c^E &\equiv p_c^E/\mathcal{P}
    \end{align}
\end{subequations}
The last two of these dimensionless groups are new relative to \citet{jenkins-jfm-2019}. The seal mobility ratio $\mathcal{M}_z^s$ compares the mobility of gas leakage to that of water leakage, and is important for interpreting the effect of the relative permeability of gas in the seal (see \S\ref{s:Mzs}). The dimensionless capillary entry pressure compares the capillary entry pressure to the characteristic pressure: Roughly, gas injection is likely to drive gas leakage if $\tilde{p}_c^E<1$ and buoyancy is likely to drive gas leakage if $\tilde{p}_c^E<(1-R_d)N_g$. We refer the reader to \citet{jenkins-jfm-2019} for a discussion of the other 8 groups.

\subsection{Model summary}

Dropping the tildes, the above scalings and dimensionless groups allows us to write our model in dimensionless form as
\begin{equation}\label{eq:GOV_NDg}
    \begin{split}
        N_{cw}(R_{cw}&\rho_{g}^{n}+R_{cf})h^n\frac{\partial{p^n}}{\partial{t}} +\rho_{g}^{n}\frac{\partial{h^n}}{\partial{t}} -\frac{\partial}{\partial{x}}\Bigg\{\rho_{g}^{n} h^n\mathcal{M}\left(\frac{\partial{p^n}}{\partial{x}}-\rho_{g}^{n}R_dN_g\frac{\partial{h^n}}{\partial{x}}\right) \\
        &+\frac{\Lambda_w^s\rho_g^{s+1}}{2s_g\rho_g^n}\frac{\partial}{\partial{x}}\bigg[(h^n)^2q_{g,z}^{s+1}\bigg]\Bigg\} =-\frac{R_A^2\Lambda_w^s}{s_g}\left( \rho_g^{s+1}q_{g,z}^{s+1} - \rho_g^{s}q_{g,z}^{s}\right) + \mathcal{I}_{g}^{n},
    \end{split}
\end{equation}
and
\begin{equation}\label{eq:GOV_NDw}
    \begin{split}
        N_{cw}(&R_{cw}+1)(1-s_gh^n)\frac{\partial{p^n}}{\partial{t}}
-s_g\frac{\partial{h^n}}{\partial{t}} -\frac{\partial}{\partial{x}}\Bigg\{(1-h^n)\bigg[\frac{\partial{p^n}}{\partial{x}} -N_g\frac{\partial{h^n}}{\partial{x}}\bigg] \\
        &+\frac{\Lambda_w^s}{6}\frac{\partial}{\partial{x}}\bigg[(1-h^n)^2(q_{w,z}^{s}+2q_{w,z}^{s+1})\bigg]\bigg\}
= -{R_A}^2\Lambda_w^s(q_{w,z}^{s+1}-q_{w,z}^{s}) + s_gR_d\mathcal{I}_w^n,
    \end{split}
\end{equation}
which enforce global conservation of mass in aquifer $n$ for $n=s=1\cdots{}N_z$. The gas density in each aquifer is related to the pressure via
\begin{equation}
    \rho_g^n(p^n) = 1 + N_{cw}R_{cf}(p^n-p^0).
\end{equation}
The aquifers are coupled by water leakage according to
\begin{equation}\label{eq:qs_system_ND}
    \begin{split}
        \frac{\Lambda_w^s}{2}(1-h^n)q_{w,z}^{s+1} +\bigg[\frac{\Lambda_w^s}{k_{rw}^\star}h^{n-1}+1+&\frac{\Lambda_w^s}{2}(1-h^n)\bigg]q_{w,z}^s \\
        &=-\bigg[p^n-p^{n-1}+N_g(h^{n-1}+b+1-h^n)\bigg],
    \end{split}
\end{equation}
for each seal $s$, $s=n=2\cdots{}N_z-1$, and by gas leakage according to
\begin{equation}\label{eq:qgzs}
    q_{g,z}^{s} = -\mathcal{R}\mathcal{M}_z^s\left(p_w^{n,B}-p_g^{n-1,T}+\rho_g^{s}R_dN_gb\right)
\end{equation}
for each seal $s$, $s=n=2\cdots{}N_z-1$, where
\begin{equation}
    p_w^{n,B} = p^n+(1-h^n)\left[N_g+\frac{\Lambda_w^s}{2}(q_{w,z}^{s+1}+q_{w,z}^s)\right]
\end{equation}
and
\begin{equation}\label{eq:pgnTdim-less}
    p_g^{n-1,T} = \frac{1}{\xi_g^{s}}\bigg[p^{n-1}-R_dN_g\rho_g^{n-1}h^{n-1} +(\xi_g^{s}-1)(p_w^{n,B}+R_dN_gb\rho_g^{s})\bigg].
\end{equation}
The capillary pressure along the underside of each seal is given by
\begin{equation}\label{eq:pcNTdim-less}
    \begin{split}
        p_c^{n-1,T} = & \left( \frac{1}{\xi_g^{s}} - \frac{1}{\xi_w^{s}}\right) p^{n-1}
        + \left( \frac{1}{\xi_w^{s}} -\frac{R_d\rho_g^{n-1}}{\xi_g^{s}}\right) N_g h^{n-1} \\
        & + \left( \frac{1}{\xi_w^{s}} - \frac{1}{\xi_g^{s}}\right) p_w^{n,B} -(1-R_d\rho_g^{s})N_g b + \left( \frac{1}{\xi_w^{s}} -\frac{R_d\rho_g^{s}}{\xi_g^{s}}\right) N_g b,
    \end{split}
\end{equation}
where
\begin{equation}\label{eq:xiw_and_xig}
    \xi_w^{s} = 1 + h^{n-1}\frac{\Lambda_w^s}{k_{rw}^{\star}} \quad\mathrm{and}\quad \xi_g^{s} = 1 + \left(\frac{\Lambda_w^s\mathcal{M}_z^s\rho_g^sh^{n-1}}{s_g\mathcal{M}\rho_g^{n-1}}\right)\mathcal{R},
\end{equation}
and the entry-pressure transition function $\mathcal{R}(p_c^{n-1,T})$ is as given in Eq.~\eqref{eq:Rpc} above. Lastly, the system is closed with boundary and initial conditions, which are written in dimensionless form as $h^n(x,t=0)=0$ and $p^n(x,t=0)=p^n(-L_x/2,t)=p^n(L_x/2,t)=p^0 -N_g[n + (n-1)b]$.

\section{Results}\label{s:results}

As evidenced by the large number of dimensionless parameters, this model incorporates a large number of physical mechanisms that will interact in complex and sometimes counter-intuitive ways. \citet{jenkins-jfm-2019} studied the coupling of fluid injection (both water and gas) with vertical and lateral pressure dissipation, the former being a direct result of water leakage. Here, having now extended the model of \citet{jenkins-jfm-2019} to allow for gas leakage subject to a capillary entry threshold, we study the coupling of gas injection with the buildup of capillary pressure and subsequent gas leakage. In doing so, we focus specifically on the likelihood of gas leakage and on the impact of gas leakage on the spatial distribution of gas at the end of injection. We plan to study the time evolution of the gas plume and the pressure field, both during and after injection, in future work.

We focus on a minimal system: Two aquifers ($N_z=2$) separated by a permeable but fine-grained seal, with gas injection into the lower aquifer $(n=1)$. We choose rock properties that are consistent with the injection of CO$_2$ into a typical target reservoir for CCS --- namely, thick sandstone aquifers and a thin mudstone seal. We choose fluid properties that are consistent with those of CO$_2$ and brine at a depth of $\sim$1~km. We assume an injection rate of $\sim$2~Mt per year distributed along a 30~km injection array, and an injection time 10~years. This scenario motivates a set of dimensional reference parameter values, from which we calculate dimensionless reference parameter values. We summarise both sets of parameters in Table~\ref{tab:params}. We use these parameter values below, except where explicitly indicated otherwise.

We solve our model numerically by discretising in space using a standard finite-volume method on a uniform grid and integrating in time using \verb+MATLAB+'s stiff ODE solver \verb+ODE15s+ \citep[][]{matlab}. At each time step, we calculate the vertical water fluxes by solving $N_x$ uncoupled linear algebraic systems of size $N_z-1$, where $N_x$ is the number of horizontal gridblocks. We also impose no vertical flow through the bottom and top of the system (\textit{i.e.}, $q_{w,z}^1=q_{w,z}^{N_z+1}=0$).

\begin{table}
    \begin{center}
        \begin{tabular}{ l c l }
            \textbf{Parameter} & \textbf{Symbol} & \textbf{Value} \\
            \\
            Number of aquifers & $N_z$ & 2 \\
            Horizontal extent & $L_x$ & 100 km \\
            Aquifer thickness & $H$ & 10 m \\
            Aquifer porosity & $\phi$ & 0.3 \\
            Aquifer permeability & $k$ & $10^{-13}$ m$^2$ \\
            Seal thickness & $b$ & 0.5 m \\
            Seal permeability & $k_s$ & $5 \times 10^{-19}$ m$^2$\\
            Seal entry pressure & $p_c^E$ & $\infty$ \\
            Rock compressibility & $c_r$ & 3.0 $\times 10^{-11}$ Pa$^{-1} $ \\
            Reference pressure & $p^0$ & 10 MPa \\
            \\
            Water viscosity & $\mu_w$ & 8 $\times 10^{-4}$ Pa$\cdot$s \\
            Water density & $\rho_w$ & 1000 kg$\cdot$m$^{-3}$ \\
            Water compressibility & $c_w$ & 4.5 $\times 10^{-10}$ Pa$^{-1}$ \\
            Saturation of water in gas region & $s_{wr}$ & 0.2 \\
            Relative permeability of water in gas region & $k_{rw}^\star$ & 10$^{-3}$ \\
            \\
            Gas viscosity & $\mu_g$ & 4 $\times 10^{-5}$ Pa$\cdot$s \\
            Gas density & $\rho_{g}^{0}$ & 700 kg$\cdot$m$^{-3}$ \\
            Gas compressibility & $c_g$ & 1.5 $\times 10^{-8}$ Pa$^{-1}$ \\
            Saturation of gas in gas region & $s_g$ & 0.8 \\
            Relative permeability of gas in gas region & $k_{rg}$ & 1 \\
            Relative permeability of gas in seals & $k_{rg}^s$ & 1 \\
            \\
            Mass injection rate & $\dot{M}$ & $2\times 10^{-3}$~kg$\cdot$s$^{-1}$$\cdot$m$^{-1}$ \\
            Injection time & $\mathcal{T}$ & 10 years \\
            \hline\hline
            Compressibility number & $N_{cw}$ & 1.21$\times 10^{-4}$ \\
            Rock-to-water compressibility ratio & $R_{cw}$ & 6.67$\times 10^{-2}$ \\
            Gas-to-water compressibility ratio & $R_{cf}$ & 33.3 \\
            Aspect ratio & $R_A$ & 18.77 \\
            Density ratio & $R_d$ & 0.7 \\
            Gravity number & $N_g$ & 0.37 \\
            Aquifer mobility ratio  & $\mathcal{M}$ & 25 \\
            Seal mobility ratio & $\mathcal{M}_z^s$ & 20 \\
            Water-leakage strength & $\Lambda_w^s$ & $10^{-4}$ \\
            Seal-to-aquifer thickness ratio & $\tilde{b}$ & 0.05 \\
            Horizontal extent & $\tilde{L}_x$ & 533 \\
        \end{tabular}
    \end{center}
    \caption{Reference parameter values. Note that the dimensionless values (below the double-line) are calculated directly from the dimensional values (above the double-line). \label{tab:params} }
\end{table}

\subsection{Evolution of capillary pressure in the absence of gas leakage ($p_c^E \to \infty$)} \label{sec:results_NGL}

Gas will begin to leak out of aquifer $n$ via seal $s+1$ when the capillary pressure along the underside of the seal exceeds the entry pressure, $p_c^{n,T}>p_c^E$. The value of $p_c^{n,T}$ is coupled to the thickness of the gas plume due to buoyancy; to the pressurisation of aquifer $n$ due to injection; to the pressurisation of aquifer $n+1$ due to vertical pressure dissipation via water leakage; and to gas leakage itself. It is therefore instructive to begin by considering the evolution of this capillary pressure in the absence of gas leakage, taking $p_c^E\to\infty$. In this limit, our model is identical to that of \citet{jenkins-jfm-2019}.

\begin{figure}
    \centering
    \includegraphics[width=\textwidth]{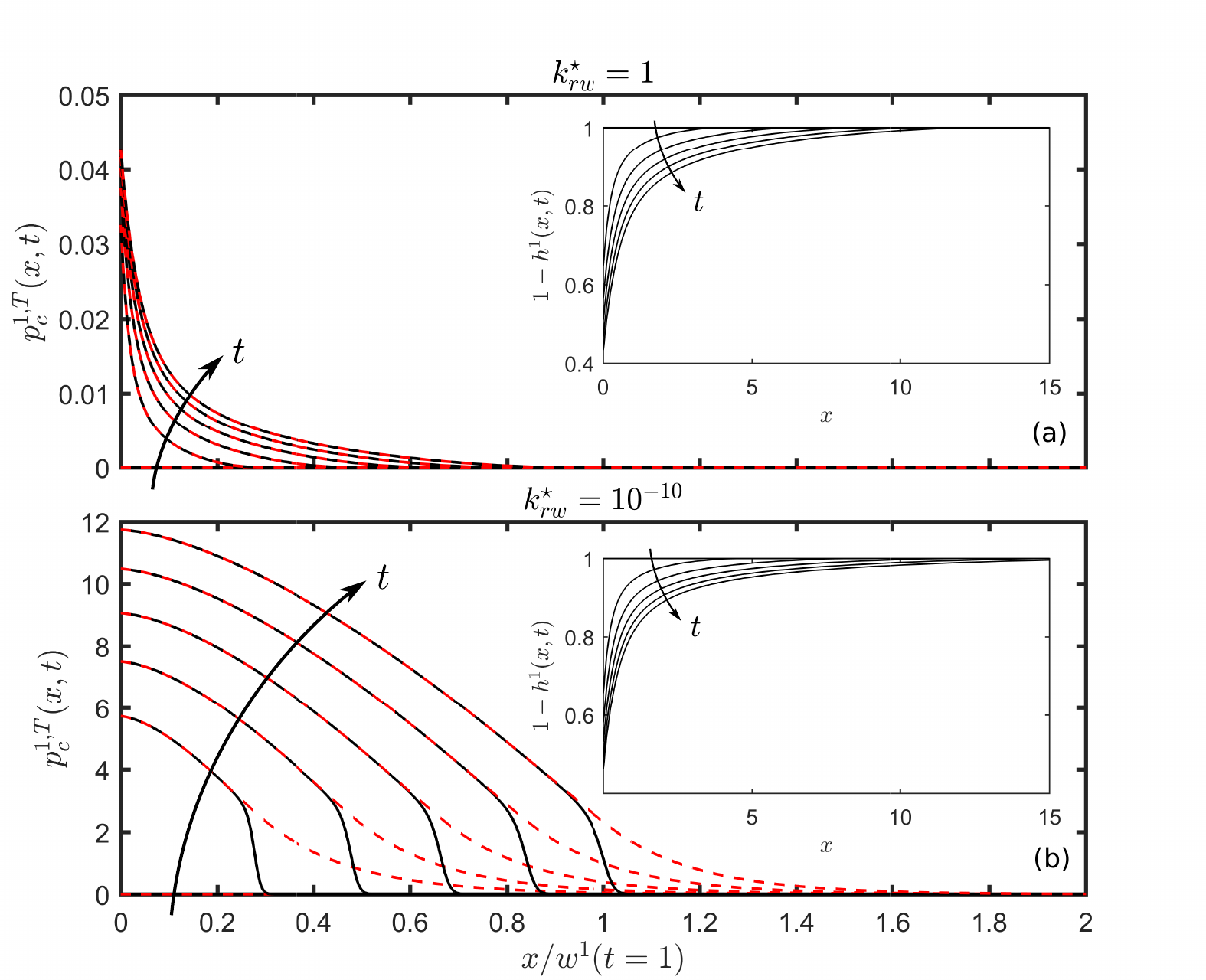}
    \caption{Gas injection into the bottom-most aquifer of a two layer system with $p_c^E=\infty$ and for (a) $k_{rw}^{\star}=1$ and (b) $k_{rw}^\star=10^{-10}$. The main plots show $p_c^{1,T}$ against $x/w(t=1)$, where $w(t)$ is the plume width, for $t=0$, $0.2$, $0.4$, $0.6$, $0.8$, and $1.0$. The insets show the plume shapes $1-h^1$ against $x$ at the same times. In the main plots, we show the actual value of $p_c^{n,T}$ as calculated from the full model (Eq.~\ref{eq:pcNTdim-less-pceinf}; solid black), as well as the approximate values appropriate to these limiting cases (Eqs.~\ref{eq:pcinf_kstar1} and \ref{eq:pcinf_kstar0} in (a) and (b), respectively; dashed red).  \label{fig:pcinf_kstar1_0} }
\end{figure}

During gas injection into aquifer $n$, a plume of gas will form, thicken, and spread with time (Figures~\ref{fig:pcinf_kstar1_0}a, inset, and \ref{fig:pcinf_kstar1_0}b, inset). As has been studied previously in some detail, the characteristic tongued shape of this plume will be controlled by several factors, including the mobility of gas relative to that of water ($\mathcal{M}$), the strength of buoyancy relative to injection ($(1-R_d)N_g$), the importance of compressibility relative to injection pressure ($N_{cw}$, $R_{cw}N_{cw}$, and $R_{cf}N_{cw}$), and the strength of vertical pressure dissipation relative to that of lateral pressure dissipation ($R_A^2\Lambda_w^s$) \citep[\textit{e.g.},][]{nordbotten2006similarity, Mathias2009approx, vilarrasa2010comp, jenkins-jfm-2019}.

The limit $p_c^E\to\infty$ implies that $\mathcal{R}=0$ and therefore that $\xi_g^s=1$ and $q_{g,z}^s=0$ for all $x$, $t$, and $s$ (Eqs.~\ref{eq:qgzs} and \ref{eq:xiw_and_xig}). As a result, the vertical pressure distribution within the gas is gas-static (Eq.~\ref{eq:pg_from_Darcy}) and $p_c^{n,T}$ is therefore given by
\begin{equation}\label{eq:pcNTdim-less-pceinf}
    p_c^{n,T} = \left(1-\frac{1}{\xi_w^{s+1}}\right) \left(p^{n}-p_w^{n+1,B}-N_gb\right) -\left(R_d\rho_g^n-\frac{1}{\xi_w^{s+1}}\right)N_{g}h^n.
\end{equation}
Note that this expression is a good approximation to $p_c^{n,T}$ even with gas leakage, because our model fundamentally assumes that the seals provide more resistance to vertical flow than the aquifers ($\Lambda_w^s\ll1$) and we expect that the gas will be much more mobile within the aquifers than the water ($\mathcal{M}\gg{}1$).

This expression above for $p_c^{n,T}$ can be further simplified by considering the value of $\xi_w^{s+1}=1+\Lambda_w^sh^n/k_{rw}^\star$ (Eq.~\ref{eq:xiw_and_xig}). Beneath and in the gas plume, vertical flow of water is resisted by two low permeabilities in series: The low relative permeability of water in the gas region ($k_{rw}^\star\ll{}1$) and the low permeability of the seals ($k_s\ll{}k$). The quantity $\Lambda_w^sh^n/k_{rw}^\star=(h^n/\lambda_w^\star)/(b/\Lambda_w^s)$ measures the ratio of the former resistance to the latter.

In the limit where the seals dominate the total resistance to upward water flow under the gas plume, $b/\Lambda_w^s\gg{}h^n/\lambda_w^\star \implies \Lambda_w^sh^n/k_{rw}^\star\ll1 \implies \xi_w\approx1$ and the capillary pressure $p_c^{n,T}$ reduces to
\begin{equation}\label{eq:pcinf_kstar1}
    p_c^{n,T}\approx(1-R_d\rho_g^n)N_gh^n \quad\mathrm{for}\quad \Lambda_w^sh^n/k_{rw}^\star\ll1,
\end{equation}
or, in dimensional terms, $p_c^{n,T}\approx(\rho_w-\rho_g^n)gh^n$. Thus, the gas pressure is gas-static from the interface upward ($p_g^{n,T}=p^n-R_dN_g\rho_g^nh^n$) and, if the gas offers negligible resistance to vertical flow of water, the water pressure is also nearly hydrostatic from the interface upward ($p_w^{n,T}\approx{}p^n-N_gh^n$). As a result, the capillary pressure at the top of the aquifer is simply given by the so-called buoyant overpressure in the gas, which mirrors the plume shape ($p_c^{n,T}\propto{}h^n$), as illustrated in Figure~\ref{fig:pcinf_kstar1_0}(a). In this limit, our entry pressure condition becomes identical to that of \citet{woods2009capillary}.

In the opposite limit, where the gas region dominates the total resistance to upward water flow under the gas plume, $b/\Lambda_w^s\ll{}h^n/\lambda_w^\star \implies \Lambda_w^sh^n/k_{rw}^\star\gg1 \implies \xi_w^{s+1}\gg{}1$ and the capillary pressure instead reduces to
\begin{equation}\label{eq:pcinf_kstar0}
    p_c^{n,T} \approx \underbrace{(p^n-R_d\rho_g^nN_gh^n)}_\textrm{gas pressure} -\underbrace{(p_w^{n+1,B}+N_gb)}_\textrm{water pressure} \quad\mathrm{for}\quad \Lambda_w^sh^n/k_{rw}^\star\gg1.
\end{equation}
Thus, the gas pressure is still gas-static from the interface upward ($p_g^{n,T}=p^n-R_dN_g\rho_g^nh^n$), but the water pressure at the top of the aquifer is now effectively hydraulically disconnected from the water pressure  at the interface by the presence of the gas. As a result, the water pressure at the top of the aquifer is given by the pressure at the bottom of the aquifer above plus the hydrostatic difference across the seal. In other words, a very small value of $k_{rw}^\star$ in the gas region (strictly, $k_{rw}^\star\ll{}\Lambda_w^sh^n$) can enable a much larger pressure difference between the gas and the water at the top of the aquifer, and therefore a much larger capillary pressure. We illustrate this in Figure~\ref{fig:pcinf_kstar1_0}(b), having changed only the value of $k_{rw}^\star$ relative to Figure~\ref{fig:pcinf_kstar1_0}(a). The shape of the gas plume is very similar to that shown in Figure~\ref{fig:pcinf_kstar1_0}(a), but now somewhat broader due to the reduction in pressure dissipation. However, the capillary pressure is now more than two orders of magnitude larger, and it remains large over a much broader region; the latter is due to the development of a plateau in the water pressure above the gas plume~\citep{jenkins-jfm-2019}. Note that the assumption that $\Lambda_w^sh^n/k_{rw}^\star\gg{}1$ must always eventually break down near the advancing nose of the plume for any nonzero $k_{rw}^\star$, since there must always exist some vanishing plume thickness at which the gas can no longer obstruct the water (\textit{i.e.}, where $h^n\not\gg{}k_{rw}^\star/\Lambda_w^s$). In other words, the former buoyancy-dominated limit of $\Lambda_w^sh^n/k_{rw}^\star\ll{}1$ must always apply sufficiently close to the edge of the gas plume.

\begin{figure} 
    \centering
    \includegraphics[width=\textwidth]{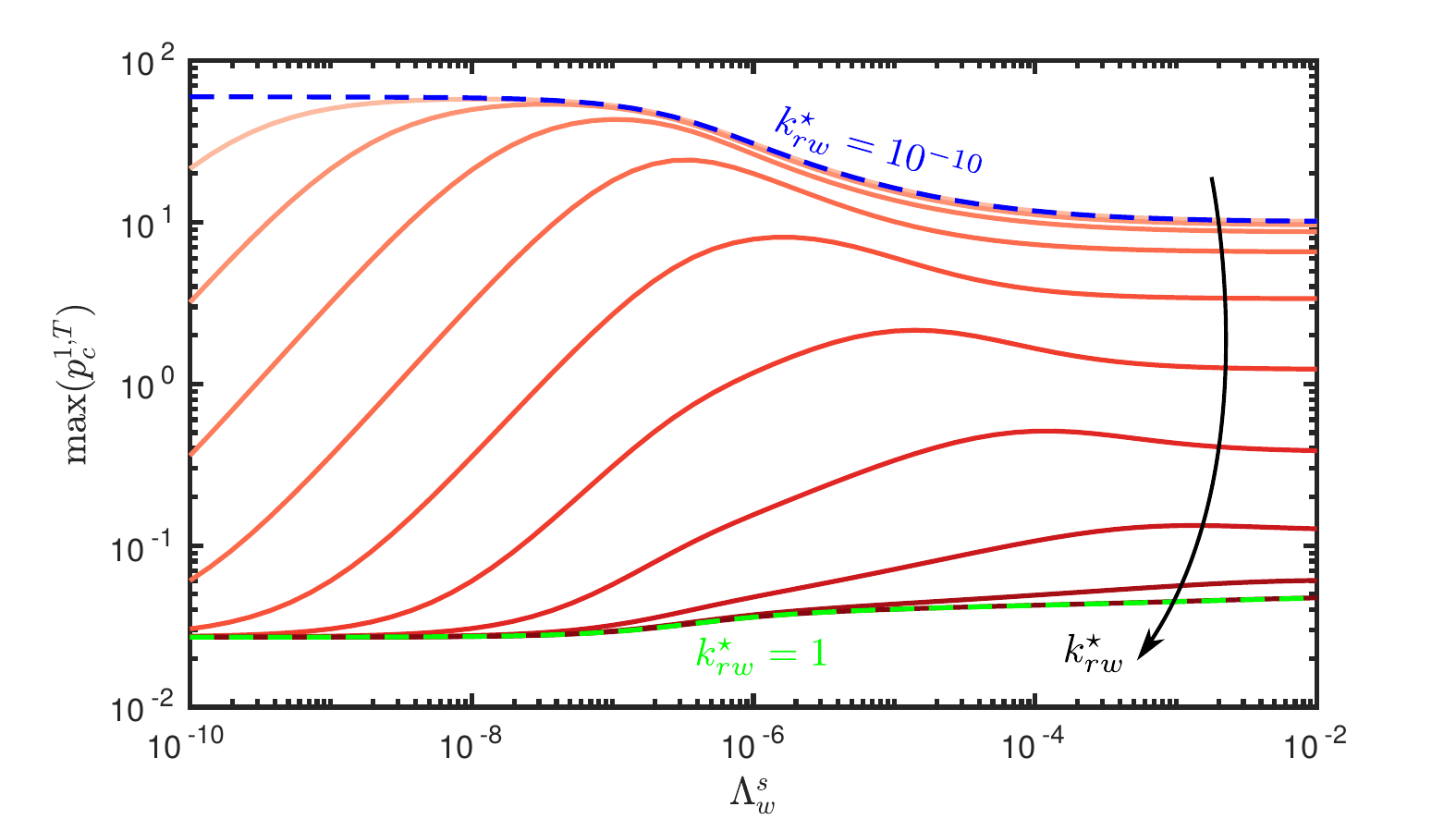}
    \caption{The maximum capillary pressure $\mathrm{max}(p_c^{1,T})=p_c^{1,T}(x=0,t=1)$ is a strong function of both $\Lambda_w^s$ and $k_{rw}^\star$. Here, we plot $\mathrm{max}(p_c^{1,T})$ against $\Lambda_w^s$ for $\log_{10}(k_{rw}^\star)=-10$, $-9$, $-8$, $-7$, $-6$, $-5$, $-4$, $-3$, $-2$, and $0$ (light to dark). We also show the prediction of Eq.~\eqref{eq:pcinf_kstar0} for $k_{rw}^\star=10^{-10}$ (dashed blue) and of Eq.~\eqref{eq:pcinf_kstar1} for $k_{rw}^\star=1$ (dashed green). Note that $\mathrm{max}(p_c^{1,T})$ decreases by two to three orders of magnitude in varying $k_{rw}^\star$ from the former limit to the latter. \label{fig:Pc_maps_NGL} }
\end{figure}

The maximum capillary pressure always occurs at the injection point ($x=0$) and at the end of injection ($t=1$), $\mathrm{max}(p_c^{n,T})=p_c^{n,T}(x=0,t=1)$, and this maximum is sensitive to both $\Lambda_w^s$ and $k_{rw}^\star$ through their respective impacts on pressure dissipation. We plot $\mathrm{max}(p_c^{n,T})$ against $\Lambda_w^s$ in Figure~\ref{fig:Pc_maps_NGL} for several different values of $k_{rw}^\star$ across its full range, highlighting the transition between the two limits discussed above ($k_{rw}^\star\approx{}1$ and $k_{rw}^\star\ll{}1$) and the orders-of-magnitude change in $\mathrm{max}(p_c^{n,T})$ as $k_{rw}^\star$ transitions between these two limits. Note also that $\mathrm{max}(p_c^{n,T})$ is weakly non-monotonic in $\Lambda_w^s$ for fixed $k_{rw}^{\star}$.

These results suggest that the buoyant overpressure may provide a substantial underestimate of the capillary pressure in the presence of vertical pressure dissipation, which has implications for the likelihood of gas leakage for typical geological parameters (see \S\ref{s:discussion}).

\subsection{Gas leakage through a uniform seal}

We now consider finite values of $p_c^E$, for which gas does eventually leak when buoyancy and/or injection is sufficiently strong. The strength and horizontal extent of gas leakage are controlled by several key parameters: $k_{rw}^\star$, as discussed above; $\Lambda_w^s$, which measures the resistance of the aquifers to water flow relative to that of seals; $\mathcal{M}_z^s$, which measures the resistance of the seals to water flow relative to their resistance to gas flow; and, of course, $p_c^E$. In the previous section, we focused on the impact of $k_{rw}^\star$ on the potential for gas leakage. In this section, we now fix $k_{rw}^\star=10^{-3}$ (the reference value) and study the roles of $\Lambda_w^s$, $\mathcal{M}_z^s$, and $p_c^E$ in determining the strength and extent of gas leakage, and the resulting distribution of gas in aquifers 1 and 2 at the end of injection.

\subsubsection{Varying $\Lambda_w^s$ and $p_c^E$ for fixed $\mathcal{M}_z^s$} \label{results:GL_fixedM}

We first fix $\mathcal{M}_z^s=20$ (the reference value) and study the roles of $\Lambda_w^s$ and $p_c^E$. To do so, we consider the total mass of gas in aquifer~2 at the end of injection. Note that the total mass of gas in aquifer $n$ at time $t$ is given by
\begin{equation}
    M_g^n(t)=\int_{-\infty }^{+\infty}\,\rho_g^{n}h^{n}\,\mathrm{d}x,
\end{equation}
and the total mass of gas in the system is given by
\begin{equation}
    M_g^\mathrm{tot}(t)=\sum_{n=1}^{N_z}\,M_g^n(t).
\end{equation}
Here, we prescribe a constant mass rate of gas injection $\mathcal{I}_g^1=2$ until time $t=1$ and it must therefore be the case that $M_g^\mathrm{tot}(t)=M_g^1(t)+M_g^2(t)=2t$. Without gas leakage, we expect at the end of injection that $M_g^1(t=1)=2$ and  $M_g^2(t=1)=0$ (\textit{i.e.}, that all of the gas is located in the injection aquifer). With gas leakage, we expect that $M_g^2(t=1)>0$ and that $M_g^1(t=1)=2-M_g^1(t=1)<2$.

\begin{figure}
    \centering
    \includegraphics[width=\textwidth]{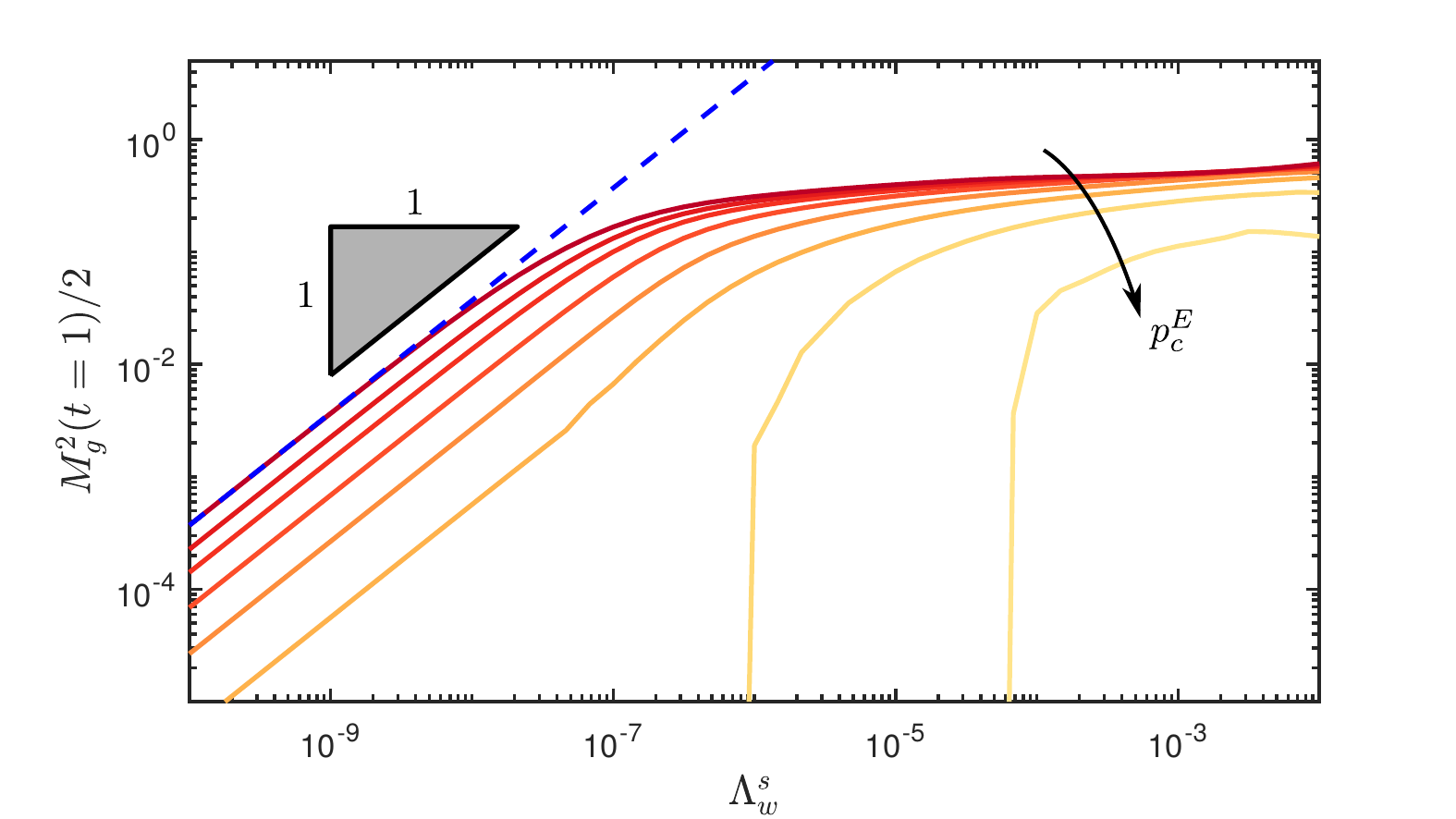}
    \caption{Mass fraction of gas in aquifer~2 at the end of injection, $M_g^2(t=1)/2$, plotted against $\Lambda_w^s$ for $\log_{10}(p_c^E)=-4$, $-3$, $-8/3$, $-7/3$, $-2$, $-5/3$, $-4/3$, and $-1$ (dark to light). The dashed blue line illustrates a slope of 1 for reference, indicating that $M_g^2(t=1)\propto\Lambda_w^s$ for small values of $\Lambda_w^s$. For the two largest values of $p_c^E$ shown here, the curves diverge downward at some critical value of $\Lambda_w^s$, below which the entry pressure is never exceeded and there is therefore no gas leakage ($M_g^2\to{}0$).  \label{fig:Mg2_vs_RLAMws_pcE} }
\end{figure}

We expect the rate of gas leakage to increase with $\Lambda_w^s$ since $q_{g,z}^1\propto{}\Lambda_w^s$. The likelihood of exceeding the entry pressure also increases strongly with $\Lambda_w^s$ for $\Lambda_w^s\ll{}k_{rw}^\star/h^1$, and is insensitive to $\Lambda_w^s$ for $\Lambda_w^s\gg{}k_{rw}^\star/h^1$ (Fig.~\ref{fig:Pc_maps_NGL}). As a result, we expect the total amount of gas leakage to increase monotonically (or nearly so) as $\Lambda_w^s$ increases. Similarly, gas leakage starts earlier and occurs over a larger horizontal extent as $p_c^E$ decreases; we therefore expect the total amount of gas leakage to increase monotonically as $p_c^E$ decreases. Figure~\ref{fig:Mg2_vs_RLAMws_pcE} illustrates this behaviour, showing that $M_g^2(t=1)$ increases monotonically as $\Lambda_w^s$ increases and as $p_c^E$ decreases.

\begin{figure}
    \centering
    \includegraphics[angle=0,width=\textwidth]{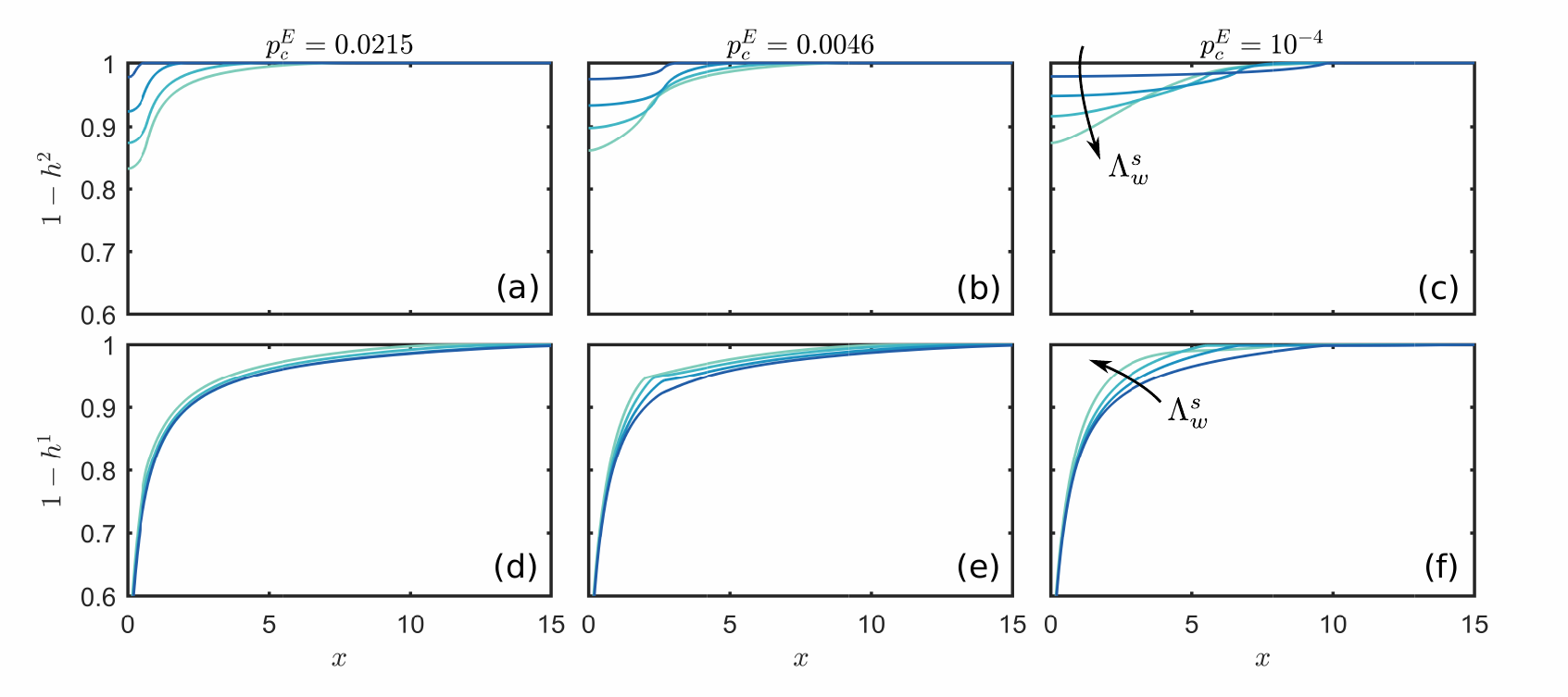}
    \caption{The shape of the gas plume in aquifer~1 (bottom row) and in aquifer~2 (top row) the end of injection, $1-h^1(x,t=1)$ and $1-h^2(x,t=1)$, respectively, for three different values of $p_c^E$ (decreasing left to right). In each panel, we show results for $\Lambda_w^s=10^{-7}$, $10^{-6}$, $10^{-5}$, and $10^{-4}$ (dark to light colors). Recall that $k_{rw}^\star=10^{-3}. $\label{fig:plumes_leak} }
\end{figure}

We now consider the shapes of the gas plumes in aquifers~1 and 2. The horizontal length over which gas leakage occurs is the horizontal length over which $p_c^{1,T}$ exceeds $p_c^E$, which depends on the horizontal distribution of $p_c^{1,T}$ and on the magnitude of $p_c^{1,T}$ relative to $p_c^E$. For larger values of $p_c^E$, leakage is localised near the injection well where $p_c^{1,T}$ is largest, as evidenced by the localised gas plume in aquifer~2 (Fig.~\ref{fig:plumes_leak}, upper left). As $p_c^E$ decreases at fixed $\Lambda_w^s$ (Fig.~\ref{fig:plumes_leak}, left to right), the gas plume in aquifer~2 grows broader as gas leakage starts earlier and occurs over a larger horizontal extent, but also thinner as the increasing amount of gas leakage begins to have a stronger impact on the gas plume in aquifer~1. For low $p_c^E$, the horizontal extent of gas leakage approaches the full width of the plume in aquifer~1, leading to a plume of the same width in aquifer~2. As $\Lambda_w^s$ increases at fixed $p_c^E$ (Fig.~\ref{fig:plumes_leak}, dark to light colors), the leakage flux increases and the width of the leaking region broadens due to the broadening of $p_c^{1,T}$ (Fig.~\ref{fig:pcinf_kstar1_0}). At high $p_c^E$, the gas plume in aquifer~2 grows thicker and broader as $\Lambda_w^s$ increases, whereas the gas plume in aquifer~1 shrinks gently over its entire extent while retaining its shape. At lower $p_c^E$, in contrast, the gas plume in aquifer~2 grows thicker and narrower as the rate of gas leakage becomes large enough to substantially decrease the extent of the gas plume in aquifer~1.

Moderate values of $p_c^E$ and moderate to high values of $\Lambda_w^s$ lead to a large leakage rate that partially but not entirely spans the gas plume in aquifer~1, the plumes in both aquifers develop a noticeable kink at the transition between the region that is losing/gaining gas and the region that is not. When gas leakage is either focused near the injection well (high $p_c^E$ and/or $\Lambda_w^s\ll{}k_{rw}^\star/h^1$) or distributed over the entire length of the gas plume in aquifer~1 (combination of low $p_c^E$ and/or $\Lambda_w^s\gg{}k_{rw}^\star/h^1$), the plumes in both aquifers are smooth.

\begin{figure}
    \centering
    \includegraphics[width=\textwidth]{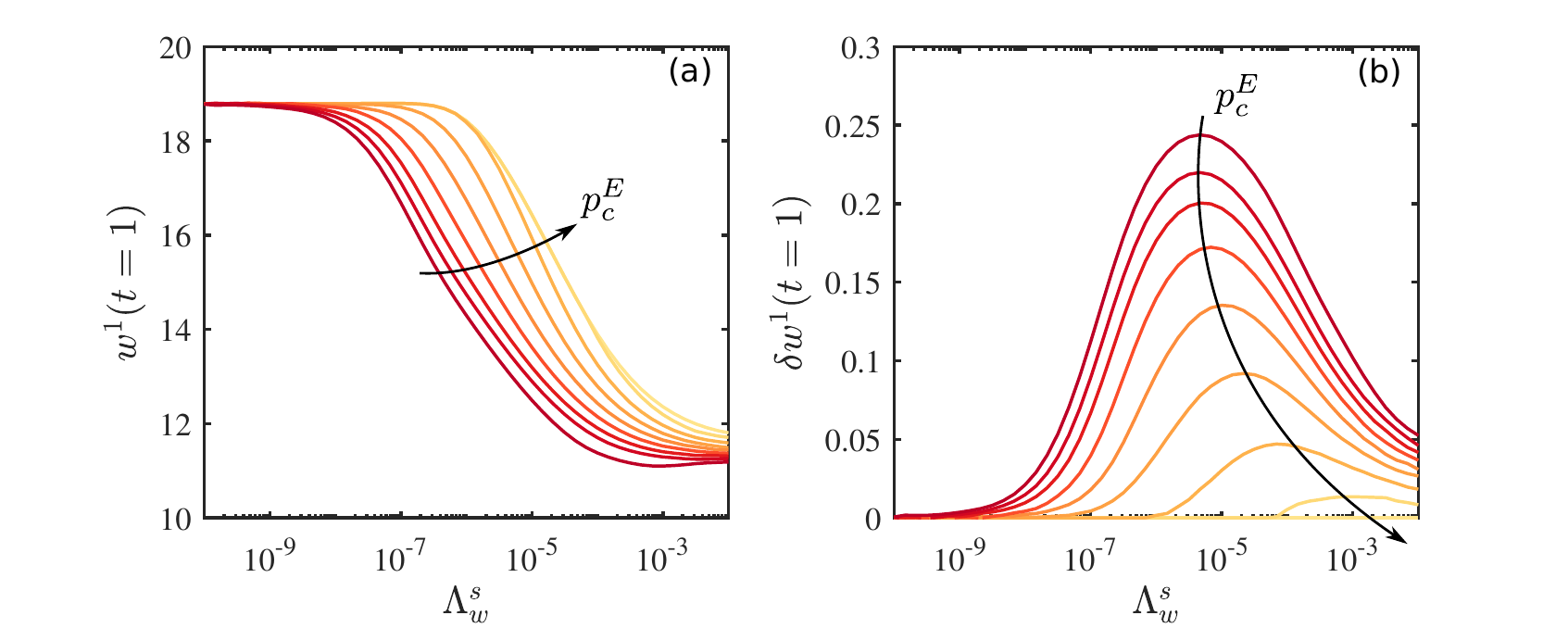}
    \caption{(a)~Plume width in aquifer~1 at the end of injection, $w^1(t=1)$, plotted against $\Lambda_w^s$ for $\log_{10}(p_c^E)=-11/3$, $-3$, $-8/3$, $-7/3$, $-2$, $-5/3$, $-4/3$, $-1$, and $0$ (dark to light). To separate the decrease in plume width due to gas leakage from the decrease in plume width due to vertical pressure dissipation, we also show (b)~the relative difference between $w^1(t=1)$ and its value for $p_c^E\to\infty$ (no gas leakage).  \label{fig:GL_RLAMws_n1} }
\end{figure}

We next quantify these plume shapes. For the gas plume in aquifer~1, we consider the plume width $w^1(t)$, where $w^n(t)$ is the distance between the injection point and the place where the thickness of the plume in aquifer~$n$ falls below an arbitrary threshold value (here, $10^{-6}$). The width of the plume in aquifer~1 at the end of injection, $w^1(t=1)$, decreases monotonically as $p_c^E$ decreases and nearly monotonically as $\Lambda_w^s$ increases (Fig.~\ref{fig:GL_RLAMws_n1}a). Recall, however, that the width of the plume in aquifer~1 decreases strongly with $\Lambda_w^s$ even in the absence of gas leakage due to increasing vertical pressure dissipation, as discussed in detail in \citet{jenkins-jfm-2019}. To separate the reduction in $w^1(t=1)$ due to gas leakage from that due to vertical pressure dissipation, we calculate the relative difference between the width for a particular value of $p_c^E$, $w^1(p_c^E)$, and the width for $p_c^E\to\infty$, $w^1(p_c^E\to\infty)$ (no gas leakage): $\delta{w^1}=[w^1(p_c^E\to\infty)-w^1(p_c^E)]/w^1(p_c^E\to\infty)$ (Fig.~\ref{fig:GL_RLAMws_n1}b). Vertical pressure dissipation dominates the reduction in $w^1(t=1)$ for very low and very high values of $\Lambda_w^s$; very little gas leakage occurs in the former case, and vertical pressure dissipation is very strong in the latter case. For intermediate values of $\Lambda_w^s$, gas leakage leads to a plume that is up to about 25~\% narrower than what would result from pressure dissipation alone.

\begin{figure}
    \centering
    \includegraphics[width=\textwidth]{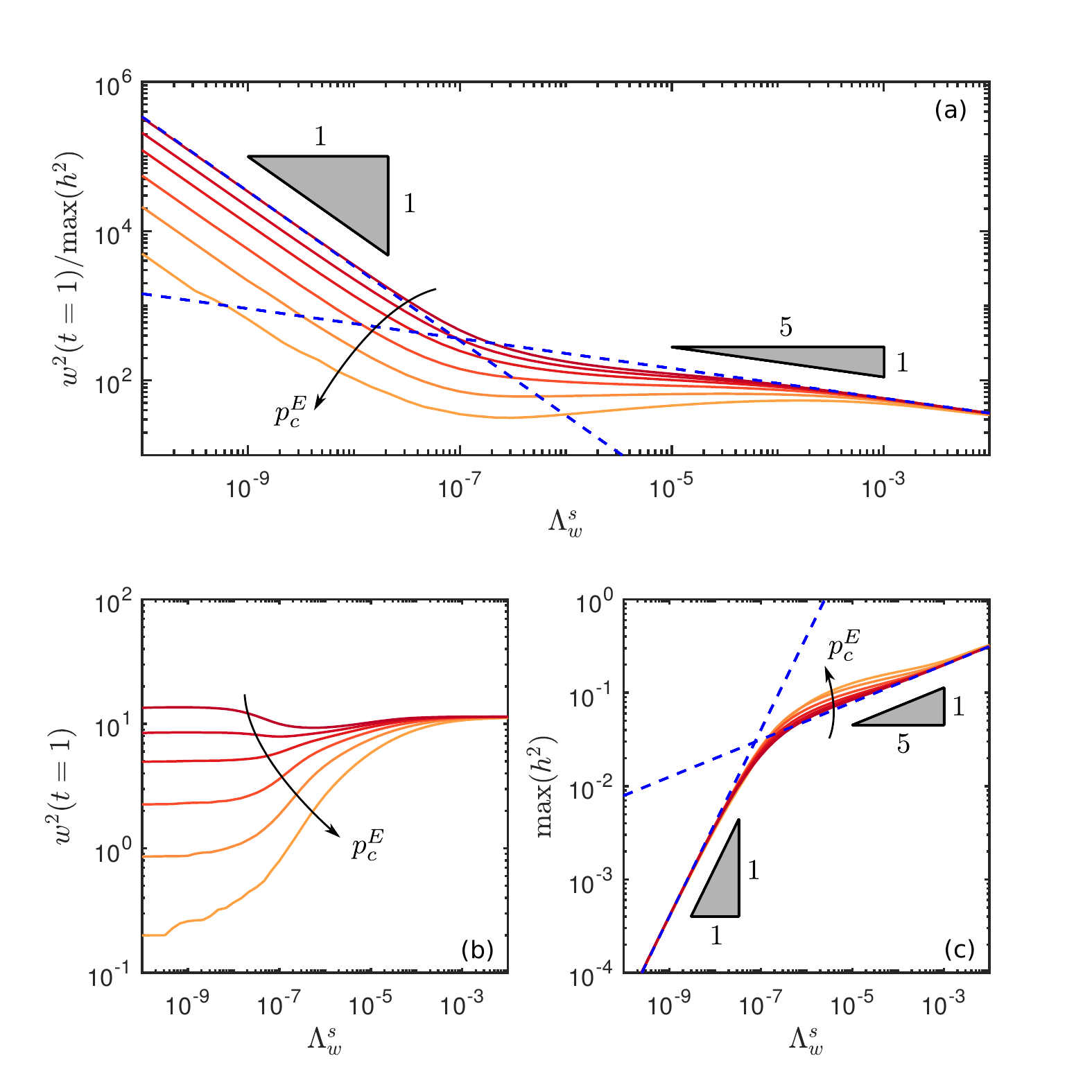}
    \caption{(a) The aspect ratio of the gas plume in aquifer~2, $\mathcal{A}^2=w^2(t=1)/\mathrm{max}(h^2)$, plotted against $\Lambda_w^s$ for $\log_{10}(p_c^E)=[ -5/3,~ -2,~ -7/3,~ -8/3,~ -9/3$ ~and~
    $-11/3]$ (dark to light), where $w^2(t=1)$ is the width of this plume at the end of injection and $\mathrm{max}(h^2)=h^2(x=0,t=1)$ is the maximum thickness of this gas plume. We also plot (b)~$w^2(t=1)$ and (c)~$\mathrm{max}(h^2)$ individually against $\Lambda_w^s$ for the same values of $p_c^E$. \label{fig:GL_RLAMws_n2} }
\end{figure}

For the gas plume in aquifer~2, the width at the end of injection $w^2(t=1)$ increases strongly and monotonically as $p_c^E$ decreases, but has a relatively weak and non-monotonic dependence on $\Lambda_w^s$ (Fig.~\ref{fig:GL_RLAMws_n2}b). The maximum thickness of the gas plume in aquifer~2, $\mathrm{max}(h^2)=h^2(x=0,t=1)$, is easier to interpret, increasing monotonically as $\Lambda_w^s$ increases and as $p_c^E$ decreases, although the latter dependence is much weaker (Fig.~\ref{fig:GL_RLAMws_n2}c). We find that $\mathrm{max}(h^2)\sim\Lambda_w^s$ for small $\Lambda_w^s$, which is expected since leakage is weak and $h^1$ and the pressure field are only mildly impacted by gas leakage. As $\Lambda_w^s$ increases, we observe a transition to $\mathrm{max}(h^2)\sim(\Lambda_w^s)^{1/5}$.

Lastly, we consider the aspect ratio of the gas plume in aquifer~2, $\mathcal{A}^2=w^2(t=1)/\mathrm{max}(h^2)$ (Fig.~\ref{fig:GL_RLAMws_n2}a). The aspect ratio increases strongly as $p_c^E$ decreases, suggesting that the increase in $w^2(t=1)$ dominates the increase in $\mathrm{max}(h^2)$---we therefore always expect a more elongated gas plume in aquifer~2 for lower $p_c^E$. With regard to $\Lambda_w^s$, however, the scaling of $\mathrm{max}(h^2)$ dominates the scaling of $\mathcal{A}^2$ because $w^2$ does not exhibit a systematic trend, so $\mathcal{A}^2\sim(\Lambda_w^s)^{-1}$ for small $\Lambda_w^s$, transitioning to $\mathcal{A}^2\sim(\Lambda_w^s)^{-1/5}$ as $\Lambda_w^s$ increases. In other words, the gas plume in aquifer~2 will always grow taller relative to its width as $\Lambda_w^s$ increases (Fig.~\ref{fig:plumes_leak}).

\subsubsection{ Effect of $\mathcal{M}_z^s$ } \label{s:Mzs}

\begin{figure}
    \centering
    \includegraphics[width=\textwidth]{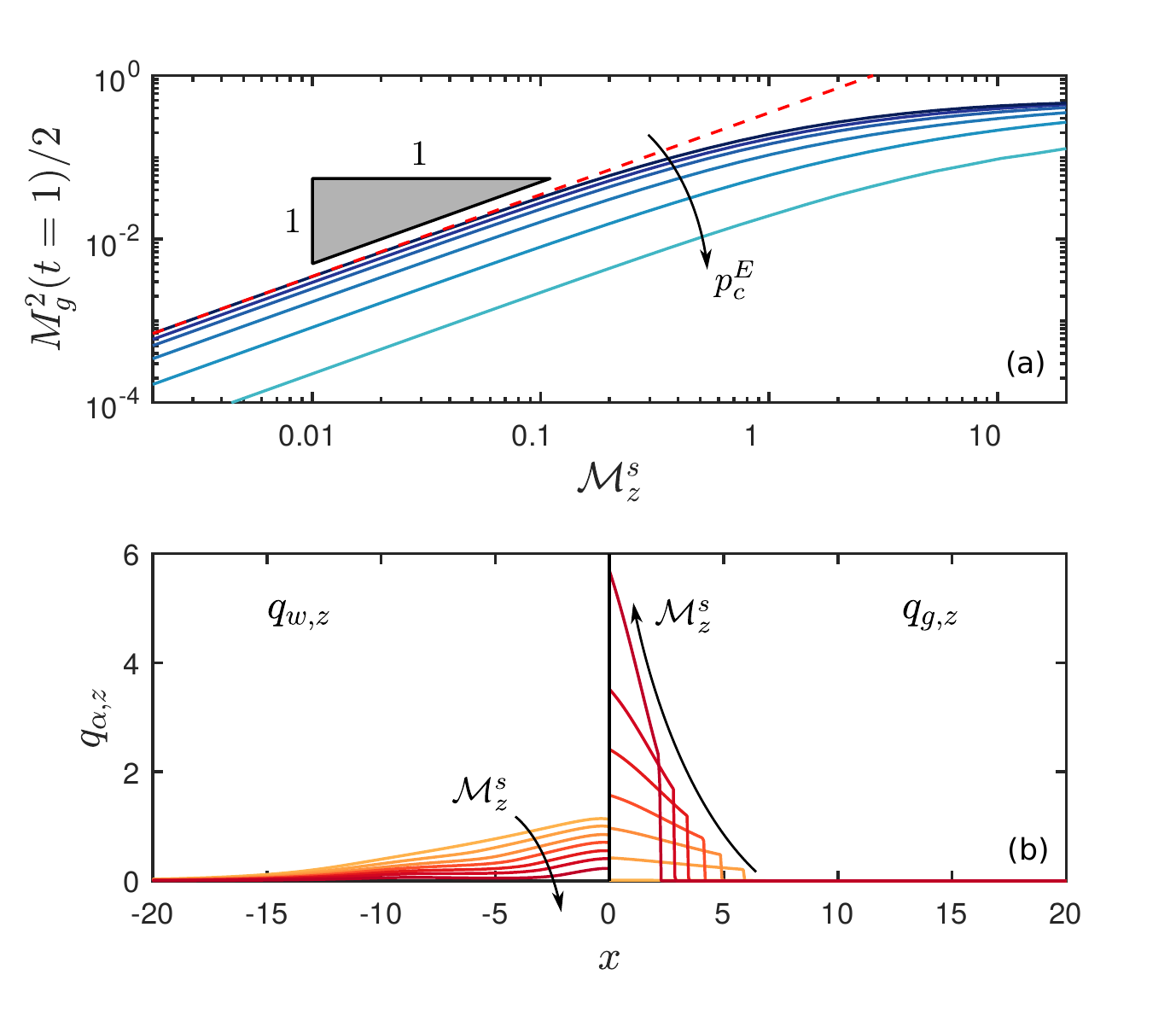}
    \caption{(a) Mass fraction of gas in the upper aquifer at the end of injection, $M_g^2(t=1)/2$, plotted against $\mathcal{M}_z^s$ for $\log_{10}(p_c^E)=-4$, $-2.8$, $-2.4$, $-2$, $-1.6$, and $-1.2$ (dark to light). (b)~Vertical leakage flux of water $q_{w,z}^s$ (left half, $x<0$) and of gas $q_{g,z}^s$ (right half, $x>0$) for a fixed entry pressure and $\mathcal{M}_z^s=0.01$, $0.39$, $1.04$, $2.00$, $3.86$, $7.46$, and $20.00$ (light to dark). \label{fig:GL_MzS} }
\end{figure}

We now briefly consider the impact of varying $\mathcal{M}_z^s$ at fixed $\Lambda_w^s=10^{-4}$ (the reference value). Recall that $\mathcal{M}_z^s\equiv (k_{rg}^s/\mu_g)/(k_{rw}^s/\mu_w)$ is the ratio of the mobility of gas in the seal to the mobility of water in the seal. Recall that $q_{g,z}^s\propto\mathcal{M}_z^s$, so we expect the rate of gas leakage and therefore also the mass of gas in aquifer~2 to increase roughly linearly with $\mathcal{M}_z^s$ for small values of $\mathcal{M}_z^s$, where gas leakage plays a weak role in the overall pressure field (Fig.~\ref{fig:GL_MzS}a). As gas leakage becomes comparable to water leakage around $\mathcal{M}_z^s\sim1$, gas leakage has an increasingly strong impact on pressure dissipation, and on the evolution of plume in aquifer~1, and $M_g^2(t=1)$ increases sublinearly as $\mathcal{M}_z^s$ increases further.

For fixed fluid viscosities, varying $\mathcal{M}_z^s$ is equivalent to varying the relative permeabilities in the seal. Physically, this can be interpreted as changing the fraction of pore volume in the seal that is available to conduct vertical gas leakage relative to vertical water leakage. In the absence of relative permeability effects, the maximum value of $\mathcal{M}_z^s$ is $\mu_w/\mu_g=20$. In this limit, the seal is more conductive to gas than to water, and the local flux of gas across the seal exceeds that of water once $p_c^{1,T}$ exceeds $p_c^E$ (Fig.~\ref{fig:GL_MzS}b). In reality, we expect $k_{rg}^s<k_{rw}^s<1$, such that relative permeability effects should generally decrease $\mathcal{M}_z^s$ relative to $k_{rg}^s=k_{rw}^s=1$. As $\mathcal{M}_z^s$ decreases, the vertical gas flux decreases and the vertical water flux increases (Fig.~\ref{fig:GL_MzS}b). For small $\mathcal{M}_z^s$, the vertical gas flux scales as $q_{g,z} \sim \mathcal{M}_z^s$.

\section{Implications for CCS}\label{s:discussion}

We now consider the implications of our results for CCS, with a particular focus on CO$_2$ injection at Sleipner---where CO$_2$ has been injected at an average rate of about $1\,\mathrm{Mt}\,\mathrm{y}^{-1}$ since 1996. Note that, for purposes of this discussion, we revert to dimensional quantities.

The entry pressure for a perfectly non-wetting phase invading a pore is approximately $p_c^E\approx{}2\gamma/r$, where $\gamma$ is the interfacial tension and $r$ is the typical pore-throat radius. \citet{chiquet2007co2} measured the CO$_2$-water interfacial tension to be $\gamma\approx{}25\,\mathrm{mNm}^{-1}$ at reservoir conditions, and \citet{kuila2013specific} measured the typical pore-throat radii of North Sea shales to be 10--$100\,\mathrm{nm}$. These values imply $p_c^E\approx{}0.5$--$5\,\mathrm{MPa}$ for a typical sealing layer at Sleipner, which is similar to the range of $2$--$5\,\mathrm{MPa}$ estimated by \citet{chadwick2003geological}.

In \S\ref{sec:results_NGL} above, we showed that the magnitude and distribution of the capillary pressure at the base of a seal is very sensitive to the relative permeability $k_{rw}^\star$ of water in the gas-saturated region, which originates in the conductivity and connectivity of residual water films there. Neglecting the resistance to water flow through this region by assuming $k_{rw}^\star=1$ implies that the water column is very well connected across the gas region, such that the water pressure is nearly hydrostatic across the gas region and the capillary pressure along the base of the seal can be very well approximated by the buoyant overpressure in the gas, $p_c\approx(\rho_w-\rho_g)gh^n$, as has been assumed previously~\citep[\textit{e.g.},][]{woods2009capillary}. To exceed an entry pressure of $p_c^E\approx{}2\,\mathrm{MPa}$ would then require a CO$_2$ column of thickness $\sim 680\,\mathrm{m}$, which is difficult to achieve. For our reference scenario, Figure~\ref{fig:pcinf_kstar1_0}(a) implies that the maximum capillary pressure for $k_{rw}^\star\approx{}1$ is $p_c\approx 0.04\mathcal{P} \approx 11\,\mathrm{kPa}$, where $\mathcal{P}\approx0.27\,\mathrm{MPa}$ is our characteristic pressure (see \S\ref{eq:scales}). This result would suggest that the capillary pressure cannot exceed the threshold for gas leakage. At Sleipner, where seismic data has been widely interpreted to show gas leakage across sealing layers, this conclusion, in addition to a vertical seismic anomaly, has led many to infer the presence of a conduit that provides a low-entry-pressure and high-permeability vertical channel for the gas leakage \citep[\textit{e.g.},][]{chadwick2003geological,chadwick20054d,bickle2007modelling}.

\begin{figure}
    \centering
    \includegraphics[width=\textwidth]{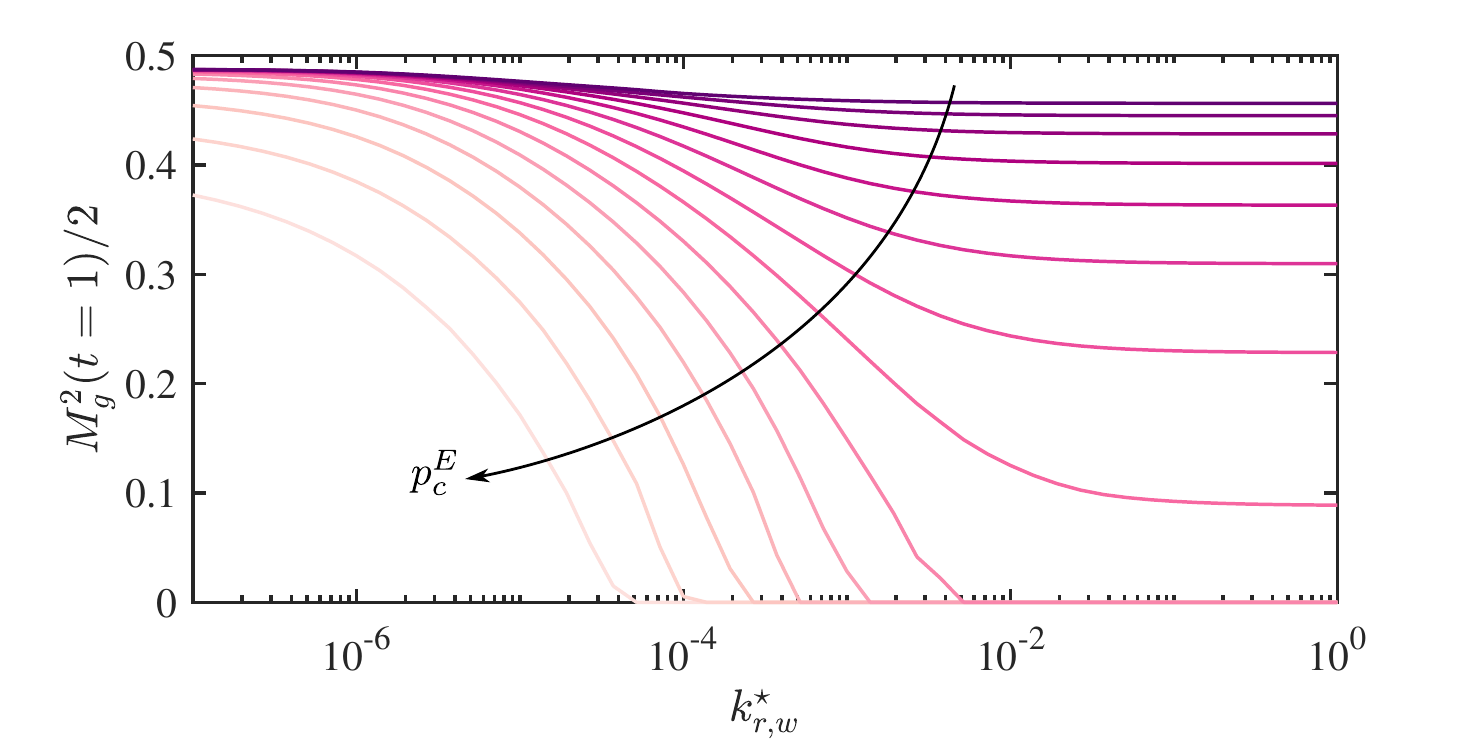}
    \caption{Mass of gas in the upper layer divided by total injected mass at the end of injection, as a function of the relative permeability of water in the gas region and plotted for $\log_{10}(p_c^E)=-3.5$, $-3.0$, $\cdots$, $-0.75$, $-0.5$, $-0.25$, and $0$ (dark to light). \label{fig:GL_kstar} }
\end{figure}

However, we demonstrated above that decreasing $k_{rw}^\star$ leads to a substantial increase in the breadth and magnitude of $p_c$. Assuming that the gas region completely obstructs vertical water flow, $k_{rw}^\star \approx 0$, implies that the water pressure is effectively disconnected across the gas region, such that the water pressure is nearly hydrostatic across the seal and the capillary pressure along the base of the seal can be very well approximated by $p_c\approx [p^1-\rho_g^ngh^n]-[p_w^{n+1,B}+\rho_w gb]$. For our reference scenario, Figure~\ref{fig:pcinf_kstar1_0}(b) implies that the maximum capillary pressure for $k_{rw}^\star\approx{}0$ is $p_c\approx 12\mathcal{P} \approx 3.2\,\mathrm{MPa}$. This capillary pressure, which is about 300 times larger than the previous estimate, is sufficiently high to suggest that gas leakage through the sealing layers at Sleipner is entirely plausible, and may occur instead of, or in addition to, gas leakage through a hypothetical conduit.

Reducing $k_{rw}^\star$ makes CO$_2$ leakage much more likely by substantially increasing the capillary pressure at the base of the seal, and would also greatly increase the amount of leakage by broadening the spatial distribution of this capillary pressure. Figure~\ref{fig:GL_kstar} shows the mass fraction of gas in the upper aquifer at the end of injection $M_g^2(t=1)/2$ (\textit{i.e.}, the fraction of injected gas that has leaked) as a function $k_{rw}^\star$ and $p_c^E$. The amount of gas leaked is very sensitive to $k_{rw}^\star$. Decreasing $k_{rw}^\star$ by 1--2 orders of magnitude can initiate leakage, or substantially amplify it. Unfortunately, the connectivity and conductivity of the water films in the gas-saturated region is poorly understood, making the appropriate value of $k_{rw}^\star$ very poorly constrained. The most likely scenario is that $k_{rw}^\star\ll{}1$, which suggests that gas leakage is much more likely than estimates based on buoyant overpressure might imply.

\section{Conclusions}\label{s:conclusion}

In this study, we derived an upscaled theoretical model for gas injection into a system of layered aquifers and seals, in which we allowed for vertical leakage of both water and gas across the intervening seals. Our model extends the framework of \citet{jenkins-jfm-2019} to include gas leakage subject to a capillary threshold. Our model is computationally efficient by virtue of being vertically integrated, which is essential for exploring the large parameter space that governs fluid migration and pressure dissipation in these systems.

After developing the model, we focused for simplicity on gas injection into a two-aquifer system. We began by studying the buildup of capillary pressure on the underside of the seal during gas injection, taking $p_c^E\to\infty$ such that no gas leakage occurs. In this limit, our model is identical to that of \citet{jenkins-jfm-2019}. We identified two end-member scenarios in the evolution of this capillary pressure, depending on the connectivity and conductivity of residual water films in the gas region, as parameterised by the relative permeability $k_{rw}^\star$ to water in that region. If the gas region provides very little resistance to vertical water flow relative to the seal ($k_{rw}^\star\gg\Lambda_w^sh_n$), then the capillary pressure along the underside of the seal is simply the phase-static pressure difference taken upward from the gas-water interface (\textit{i.e.}, the buoyant overpressure at the top of the gas region) \citep{woods2009capillary}. As a result, the spatial distribution of the capillary pressure mirrors the shape of the gas plume: It has a sharp maximum at the injection well that declines rapidly towards the thin tongue. If the gas region instead provides the principal resistance to vertical water flow relative to the seal ($k_{rw}^\star\ll\Lambda_w^sh_n$), then the water pressure at the base of the seal is effectively disconnected from the water pressure at the gas-water interface and the capillary pressure is instead related to the hydrostatic pressure measured downward through the seal from the aquifer above. This limit leads to capillary pressures that are much larger (by more than two orders of magnitude for the scenarios studied here) over a much wider region in space. We showed that this latter limit, in particular, may enable CO$_2$ leakage across interbedded shales at Sleipner. 

Having established the key role of $k_{rw}^\star$ in the buildup of capillary pressure, we subsequently studied the importance of several key parameters on vertical gas leakage. We showed, as described previously by \citet{woods2009capillary} in the context of a single-layer model, that the capillary entry pressure sets the horizontal length scale over which gas leakage occurs: Gas leakage starts earlier and occurs over a broader region as $p_c^E$ decreases. We demonstrated that increasing the conductivity of the seals relative to the aquifers (increasing $\Lambda_w^s$) increases the rate and therefore the total amount of gas leakage, as does increasing the mobility of gas in the seals relative to water (increasing $\mathcal{M}_z^s$). For small $\Lambda_w^s$ and/or small $\mathcal{M}_z^s$, the global pressure field is dominated by the water and the total amount of gas leakage increases linearly with $\Lambda_w^s$ and $\mathcal{M}_z^s$. As these parameters grow larger, gas leakage plays an increasingly important role in the pressure field and begins to suppress further growth in the leakage rate. We also showed that reducing $p_c^E$ increases the aspect ratio of the leaked gas plume (in aquifer~2), whereas increasing $\Lambda_w^s$ decreases this aspect ratio.

Our results highlight the fact that vertical pressure dissipation via water leakage establishes a global pressure field that then plays a central role in the complex interactions that control gas injection, migration, and leakage in layered aquifers, even when the seals have a very low permeability relative to the aquifers (note the range of $\Lambda_w^s$ values considered here). One consequence of these interactions is the unexpectedly important role of $k_{rw}^\star$ for pressure dissipation, plume shape, and capillary pressure, as discussed here and in \citet{jenkins-jfm-2019}; the value of this quantity is very poorly constrained, but likely to be small.

Our results suggest that aquifers and other subsurface reservoirs should not be considered in isolation when any pressurisation is expected, even for reduced-order models and even when the fluids of interest are expected to remain contained (\textit{e.g.}, CO$_2$, oil, or gas). This idea is not a new one among hydrologists and hydrogeologists~\citep[\textit{e.g.},][]{hunt-wrr-1985}, but it would appear that its implications for CCS and hydrocarbon production have not been fully appreciated.

It is straightforward to extend our model to include topography, lithological heterogeneity (\textit{e.g.}, conduits, faults or fractures), and trapping mechanisms (\textit{e.g.}, residual trapping and dissolution trapping). These topics are the subject of ongoing work.


This study was partially funded by the Natural Environment Research Council (NERC) Centre for Doctoral Training (CDT) in Oil \& Gas (grant no. NE/M00578X/1) and funding from Shell International B.V.



%

\end{document}